\documentclass[12pt,useAMS,usenatbib,article]{mn2e}
\usepackage{graphicx}
\usepackage{epsfig}
\usepackage{amssymb,amsmath}	
\usepackage{mathrsfs}
\usepackage[utf8x]{inputenc}
\usepackage[usenames]{color}
\usepackage{color}

\PrerenderUnicode{áéíóúñ}

\title[Cooling-induced structure in collapsars]{Cooling-induced structure formation and evolution in collapsars}

\author[A. Batta \& W.H. Lee]{Aldo Batta \& William H. Lee
$^{1}$\thanks{E-mail:abatta@astro.unam.mx, wlee@astro.unam.mx} \\
$^{1}$Instituto de Astronom\'\i a, Universidad Nacional Aut\'onoma de M\'exico, Apdo. postal 70-264 Ciudad Universitaria, D.F., M\'exico
}

\begin{document}

\date{Accepted . Received ; in original form }

\pagerange{\pageref{firstpage}--\pageref{lastpage}} \pubyear{}

\maketitle

\label{firstpage}

\begin{abstract}

The collapse of massive rotating stellar cores and the associated accretion onto the newborn compact object is thought to power long gamma ray bursts (GRBs). The physical scale and dynamics of the accretion disk are initially set by the angular momentum distribution in the progenitor, and the physical conditions make neutrino emission the main cooling agent in the flow. The formation and evolution of structure in these disks is potentially very relevant for the energy release and its time variability, which ultimately imprint on the observed GRB properties. To begin to characterize these, taking into account the three dimensional nature of the problem, we have carried out an initial set of calculations of the collapse of rotating polytropic cores in three dimensions, making use of a pseudo-relativistic potential and a simplified cooling prescription. We focus on the effects of self gravity and cooling on the overall morphology and evolution of the flow for a given rotation rate in the context of the 
collapsar model. For the typical cooling times expected in such a scenario we observe the appearance of strong instabilities on the cooling time scale following disk formation, which modulate the properties of the flow. Such instabilities, and the interaction they produce between the disk and the central object lead to significant variability in the obtained mass accretion and energy loss rates, which will likely translate into variations in the power of the relativistic outflow that ultimately results in a GRB .

\end{abstract}


\section{Introduction}\label{Intro}

Gamma Ray Bursts (GRBs) are bright flashes of radiation with a spectral energy peaking in the $\gamma$-ray band (Fishman \& Meegan 1995). Their duration ranges from fractions of a second to a few
minutes and produce directed fluxes of relativistic material with kinetic luminosities exceeding
$10^{53}\mbox{ erg s}^{-1}$. In order to produce such high luminosities, and observed milli-second variability in flux, it is common and reasonable to invoke the presence of a compact object (such as
a neutron star, NS, or a black hole, BH) as part of the GRB production mechanism at the level of the central engine. To date, despite the lack of definitive direct evidence, it is generally thought that GRBs are the result of cataclysmic events involving NSs (perhaps magnetized) or BHs and violent, so-called hypercritical accretion which produces a violent episode of energy release that is subsequently transformed into electromagnetic radiation (see Piran 2004; Nakar 2007; Lee \& Ramirez-Ruiz 2007; Gehrels et al. 2009 for comprehensive reviews). 

From GRB afterglow observations that cover the range from X-rays to radio, it was possible to locate
the origin of this sources at cosmological distances (see van Paradijs et al. 2000 for an initial review). Short GRBs are typically located at $z\lesssim$ 1 meanwhile, long GRBs are located at $z\simeq1-5$ or more, and compete with quasars for the most distant objects observed. On the whole, the hosts of SGRBs,
and by extension the progenitors, are not drawn from the same parent population of long GRBs.
SGRBs appear to be more diffusely positioned around galaxies, and their associated hosts contain
a generally older population of stars (Lee \& Ramirez-Ruiz 2007, Berger 2011). Observations also show that
long GRBs can sometimes be associated with a SN (with no H lines, i.e. type Ib or Ic) taking
place at the same time and at the same place.  The observation of GRB980425 in conjunction with
one of the most unusual SNe ever seen (SN1998bw, Galama et al. 1998b) was the first of this
kind, but not the last. Further, the host properties, and the locations of the bursts within them, provide strong evidence that they are related to vigorous star formation, strengthening the link with the death of massive stars (Fruchter et al. 2006, Levesque 2013). The review by Woosley \& Bloom (2006) shows the existing evidence for the link of long GRB at low redshift with type Ic SNe, and the progenitor mechanisms currently explored. 

Woosley (1993) proposed the collapsar model to explain the formation of a GRB, from a pre-supernova star in which the shutdown of nuclear reactions in the core leads to collapse and the formation of a black hole, rather than a neutron star. The accretion of the infalling material would form an accretion disk, releasing energy to power the burst. Two main variants on this model may occur. In the first, the Fe core  is massive enough to induce a direct collapse of the core into a
BH, while in the second an intermediate stage produces a proto-neutron star first, which later collapses after enough matter has been accreted onto its surface (typically this would take a few seconds). 

Two key ingredients, presumably associated to the progenitor, make this a relatively rare occurrence (as they must, considering that the rate of core-collapse SNe far exceeds the observed GRB rate): rotation, and the lack of a hydrogen envelope. The first is necessary in order to ensure that a large fraction of the available energy is released in a disk close to the black hole through accretion, rather than simply be swallowed whole by the BH (in something akin to Bondi accretion). Evolutionary models for rapidly rotating stars (Woosley \& Heger 2006) show that the core is able to retain enough angular momentum to make this a possibility depending on the mass loss history and the presence (or lack) of magnetic fields coupling the envelope to the inner regions. The second is required in order for the relativistic jet that is launched at the center of the star to perforate it, break out, and eventually lead to high energy emission far from the stellar surface, which we observe as a GRB. The envelope may either 
have been lost through interaction with a binary companion, or, if efficient mixing occurs throughout the star, the giant phase may be avoided altogether keeping the radius relatively small (Yoon \& Langer 2005, Woosley \& Heger 2006). Once a centrifugally supported disk forms, the temperature  will be high enough that neutrino emission becomes the main cooling mechanism, as pointed out in the context of supernovae by Chevalier (1989) and Houck \& Chevalier (1991), allowing accretion to proceed at extremely high rates with the attending energy release. In principle the burst itself may be powered by a combination of neutrinos themselves, or magnetic mechanisms that tap the rotation in the disk and/or the black hole. MacFadyen \& Woosley (1999) carried out a the first detailed numerical study of the collapsar, and further explored jet production, propagation and breakout from the star for a variety of configurations (MacFadyen, Woosley \& Heger 2001, Zhang, Woosley \& MacFadyen 2003). One possibility is that 
the explosion eventually does launch the stellar envelope outward and produces an extremely energetic event, leading to the observed hypernovae.

An important point is that the amount of angular momentum in the star is crucial for GRB production (Lee \& Ramirez-Ruiz 2006). Too much of it results in an accretion disk that forms far from the
black hole. The temperatures and densities are then not high enough for efficient cooling through neutrinos, and hypercritical accretion cannot proceed. Too little of it leads to quasi-spherical accretion, where the mass accretion rate can be extremely large but with near zero efficiency for the conversion of gravitational binding energy into thermal energy through shocks (and subsequently radiation). A critical assessment of the outcome of collapsing cores has recently been given by Dessart et al. (2012). 

Much effort has been applied to explore the behavior of these systems in two dimensions, assuming azimuthal symmetry (e.g., Proga et al. 2003, Fujimoto et al. 2006, Nagataki et al. 2007, Nagataki 2009, L\'{o}pez-C\'{a}mara et al. 2009, 2010, Sekiguchi \& Shibata 2011). However, relatively little work has been done in three dimensions, thus neglecting the potentially important role of the self--gravity of the infalling gas, and generic instabilities in 3D. The first study we are aware of was that carried out by Rockefeller, Fryer \& Li (2006), and based on a $60M_{\odot}$ rotating Pop III star. They observed the formation of instabilities within the disk, mainly as spiral waves, which contribute to angular momentum transport. More recently, Taylor, Miller \& Podsiadlowski (2010), considered a very rapidly rotating progenitor, formed from the merger  of two He stars. They also saw the development of  instabilities in the disk and based on the observed energy release concluded that some of their models are 
indeed capable of producing a GRB. We note that all of these models consider that the black hole is motionless at the origin of coordinates, which is a good approximation if the disk mass is negligible (which may not be the case) or if the infalling layers do not show strong asymmetries in their mass distribution. 

In this paper we focus on the dynamical effects in three dimensions, coming essentially from the self-gravity of the infalling gas and cooling, which occur during the initial stages of the collapse of a rotating polytropic envelope onto a central black hole in the context of the collapsar model. Rather than make use of a very complex equation of state, we consider generic solutions with simplified cooling (assumed to be through neutrinos). In particular, we pay special attention to the comparison between the cooling and dynamical time scales in order to gauge their effect on the formation of structure and the evolution of the mass accretion and energy losses. 
Section~2 deals with our setup and input physics, section~3 presents our results, and our conclusions are given in section~4.


\section{Initial Conditions and Input Physics}\label{Initial conditions}

In the context of the collapsar model, we studied the collapse and accretion of $2.5M_{\odot}$
rotating polytropic envelopes with adiabatic index $\gamma=5/3$ onto a $2M_{\odot}$ BH fixed at the center of the mass distribution. In this section we will present the relevant details of the initial conditions
and physical processes taken into account in our simulations which were carried out with GADGET-2 (Springel 2005; Springel, Yoshida \& White 2001), which we have modified in order to include the necessary physics to account for accretion and cooling in this context. 

\subsection{The Polytropic Envelope and its Characteristics}

The polytropic envelope was constructed by solving the Lane-Emden equation for hydrostatic equilibrium (see Shapiro \& Teukolsky 1983), from which we obtained radial profiles for the density and internal energy of $4.5M_{\odot}$ polytropic stars with central density $\rho_{c}=2.53\times10^{9}\mbox{ g cm}^{-3}$ and $\gamma=5/3$. From the density profile we mapped a 3D particle distribution with an accept/reject MonteCarlo procedure. To account for the BH, we removed the innermost $2M_{\odot}$ from this 3D polytropic star and concentrated it in  a $2M_{\odot}$ sink particle placed at the center of the distribution, leaving the remaining mass unaffected, now with inner radius $r_{int}$. In Table~\ref{table1} we show the total mass of the system $M_{s}$, as well as the initial outer and inner radii of the polytropic envelope ($R_{s}$ and $r_{int}$ respectively). From these quantities we can obtain a characteristic time scale $t_{dyn}=\sqrt{R_{s}^{3}/GM_{s}}$ from which all physical units are scaled in the code.

\begin{table}
 \begin{center}
  \begin{tabular}[!h]{c|l}
\hline
  System parameter & value\\\hline
$M_{s}$ &    $4.5\ M_{\odot}$\\
$M_{env}$  &   $2.5\ M_{\odot}$ \\
$R_{s}$ &      $1715.7$ km\\
$r_{int}$ &    $844.69$ km\\
$t_{dyn}=\sqrt{R^{3}/GM_{s}}$ &    $0.0919$ s
\\\hline
\end{tabular}
\end{center}
\caption{Envelope and total system masses, along with the characteristic time scale $t_{dyn}$
and the outer and inner radius of the envelope.}
\label{table1}
\end{table}

\begin{figure}
\centering
    \includegraphics[height=2.35in]{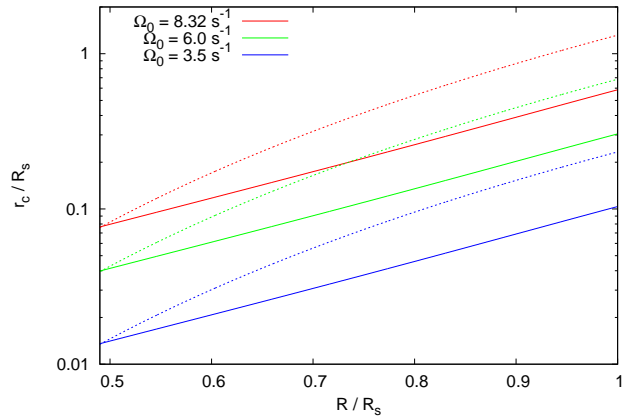}
  \caption{Circularization radius $r_{c}$ in the equatorial plane for our $2.5\ M_{\odot}$
polytropic envelope as a function of the cylindrical distance $R/R_{s}$. Each color
represents a different angular velocity $\Omega_{0}$. Dotted lines only consider the BH
mass, and solid lines consider the BH mass and an the spherical envelope mass contained at
$r_{i}$.}
\label{rcirc}
\end{figure}

\begin{figure}
  \centering
  \includegraphics[width=3.25in]{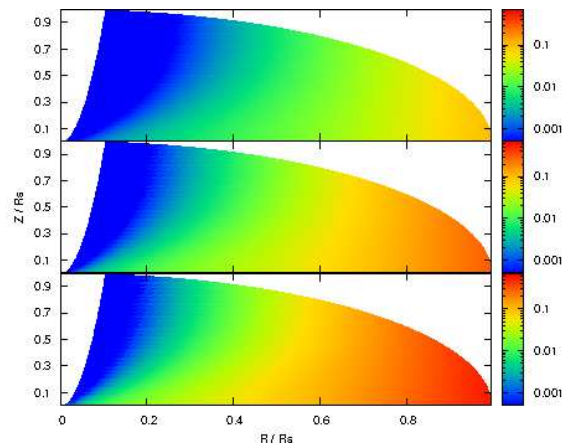}
  \caption{Circularization radius $r_{c}$ for our $4.5\ M_{\odot}$ polytropic star as a function
of the cylindrical distance $R/R_{s}$ ($X$ axis) and $Z/R_{s}$ ($Y$ axis) for
three different angular velocities ($\Omega_{0}=8.32,\ 6.0,\ 3.5 \mbox{ s}^{-1}$ from bottom to
top). This map assumes a spherical mass distribution $M(r_{i})$.}
\label{rcirc2}
\end{figure}

For the distribution of angular momentum in the star, of great relevance for the collapsar model, we have assumed rigid body rotation, and assigned a constant angular velocity $\Omega_{0}$, defined in terms of the circularization radius $r_{c}$ the infalling gas would have in a Keplerian orbit around the BH. This circularization radius is estimated assuming a Newtonian gravitational potential (a pseudo-Newtonian potential like Paczynski \& Wiita's would result in a smaller $r_{c}$, but differing in less than $\sim3\%$ from the Newtonian one at the values involved here), a negligible contribution
from gas pressure, and conservation of angular momentum for the envelope material at its initial
radius $r_{i} = (R_{i}^{2} + z^{2})^{1/2}$, where $R_{i}$ is the cylindrical radius. Thus
$r_{c}$ can be expressed as:
\begin{equation}
 r_{c}=\Omega_{0}^{2}R_{i}^{4}/GM(r_{i})
\label{Rcirceq}
\end{equation}
where $M(r_{i})$ is the mass contained inside $r_{i}$. We have assumed spherical symmetry in the mass distribution, so the gravitational pull only depends on the radius $r_{i}$. The former consideration implies that $r_{c}$ would be underestimated. Figure~\ref{rcirc} shows $r_{c}$ as a function of the initial cylindrical radius $R_{i}$ for our $2.5\ M_{\odot}$ polytropic envelopes at the equatorial plane. Each color represents a different constant angular velocity, the dotted lines show the circularization radius $r_{c}$ neglecting gravitational interaction with the rest of the envelope, whereas solid lines represent $r_{c}$ considering both the BH and the envelope gravitational interaction (with spherical
symmetry). In reality, given that rotation and gravitational collapse would induce a symmetry
breakup on the envelope, $r_{c}$ would take values somewhere in-between the dotted and solid
lines for a given angular velocity $\Omega_{0}$. For simplicity, from now on, we will only
focus on the circularization radius $r_{c}$ for the innermost envelope material at $r_{int}$, and so $r_{c}$ will give an approximate position for the inner edge of the disk with respect to the BH. Figure~\ref{rcirc2} shows the color-coded circularization radius in a meridional slice. Clearly material close to the equator has the greatest rotation rate and will circularize at a larger radius. Close to the rotation axis angular momentum goes to zero and matter will essentially free fall into the BH.

\begin{figure}
  \centering
 \includegraphics[width=2.35in,angle=-90]{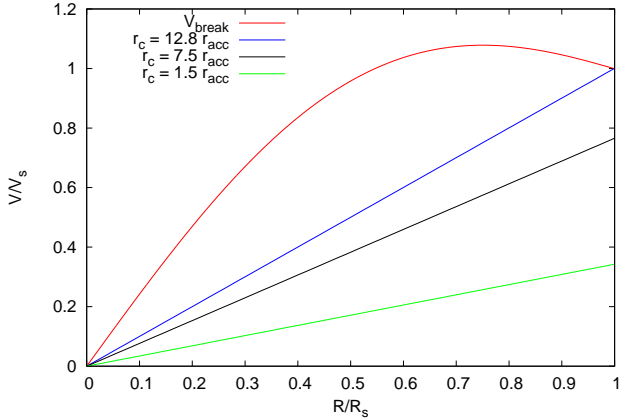}
  \label{t_nu}
  \caption{Normalized breakup velocity $V_{bk}$ as a function of radius for our $4.5\ M_{\odot}$
polytropic star (red line), and rotation velocities on the equatorial plane for rigid body
rotation with circularization radius ($r_{c}\propto \Omega_{0}^{2}$)
$12.5\ r_{acc}$ (blue line), $6\ r_{acc}$ (black line) and $r_{acc}$ (green line)).}
\label{VcbreakPol}
\end{figure}

With the purpose of choosing an adequate value for $\Omega_{0}$, we first obtained from the
initial spherical profiles, an estimation of the breakup rotational velocity for such polytropic
envelope. Then, assuming rigid body rotation, we compared the rotational velocities obtained for
an angular velocity $\Omega_{0}$ with the breakup rotational velocity for the polytrope. Figure~\ref{VcbreakPol} shows the calculated breakup rotational velocity for such a polytropic star (red line) and rotational velocities in the equatorial plane for different values of the circularization radius
$r_{c}$ ($r_{c}\propto\Omega_{0}^{2}$). Rotational velocities below the breakup velocity curve
ensure formation of a disk around the BH at $r_{c}$ ($r_{c}$ for the innermost material). This
estimation does not take into account asymmetric effects on the mass distribution due to
rotation, given that the envelope should be rotating from the beginning and therefore, it could
depart from spherical symmetry. Figure~\ref{VcbreakPol} shows that in order to obtain rotating velocities below breakup, we have to consider a circularization radius $r_{c}$ smaller than $\simeq12.8\ r_{acc}$,
where $r_{acc}$ is the accretion radius (its value is given in the BH physics which  follows).
All simulations were made using $r_{c}=7.49\ r_{acc}$ and thus correspond to maximal disks, placing as much matter as possible in the disk while avoiding centrifugal mass loss. The estimates for energy release which we obtain should be considered accordingly.

The previous setup ensures the formation of an accretion disk outside of the region of the innermost circular stable orbit around a Schwarzschild BH given by $r_{isco}=3r_{g}=6GM_{BH}/c^{2}$. Gas orbiting
the BH at $r<r_{isco}$ would fall inevitably onto it, no matter its rotational velocity.

In order to have a handle on the accuracy of the simulations, we carried out convergence tests of the collapse and accretion of such rotating polytropic envelopes onto an accreting sink particle, using 50,000, 500,000 and 5,000,000 SPH particles of equal mass. After substantial testing, we have found no relevant difference in the accretion rates observed, or in the general properties of the accretion disk, when using more than 500,000 SPH particles, and so all runs described in this paper have this initial resolution. Moreover, given that the snapshots obtained for such resolution had a reasonable size ($\lesssim 30 MB$), we were able to achieve a very good time resolution between snapshots $\sim 1/20 t_{dyn}$ which proved to be helpful when looking for structure formation at the accretion disk. 

Given the nature of our work, it is of great importance to ensure that the accretion disk has enough particle resolution to observe the formation of structures such as clumps or spiral arms. As shown by Bate \& Burkert (1997), in order to properly represent fragmentation in an SPH simulation, the Jeans mass in the disk must be greater than the minimum resolvable mass $M_{min}$ given by:
\begin{equation}
M_{min} = 2M_{tot}( N_{neigh}/ N_{tot} ),
\end{equation}
where $M_{tot}$, $N_{tot}$ and $N_{neigh}$ are the total gas mass, the total number of particles and the SPH number of neighbors, respectively. For all of our simulations we found that, at all times, the Jean's mass $M_{J}$ at radii $R< 0.9 R_{p}$ was at least one order of magnitude above the resolution mass of the simulation. Meanwhile, for $R > 0.9 R_{p}$ this condition was not always satisfied. It is therefore safe to say that for $R \lesssim 0.9 R_{p}$, where more than 90\% of the disk's mass is contained, the formation of structure is properly resolved. 

\subsection{Black Hole Physics}

We consider the formation of a Schwarzschild BH at the
center of the distribution from the innermost $2M_{\odot}$ of the initial $4.5M_{\odot}$
polytropic star. In general relativity (GR), the Schwarszchild solution for a non rotating,
neutral spherical BH of mass $M_{BH}$ implies the definition of the gravitational radius
$r_{g}=2GM_{BH}/c^{2}$. This radius represents the event horizon, and therefore there cannot be
static observers within $r<r_{g}$. When studying the motion of a test particle in the
Schwarzschild metric, and considering a circular orbit at radius $r$ around the BH, a relationship  emerges  between the orbital angular momentum $L$ and the radius $r$, from which the innermost
stable circular orbit can be obtained as $r_{isco}=3r_{g}$. At $r>3r_{g}$ there can be stable
circular orbits, and at $r<3r_{g}$ they are all unstable. Given that at distances $r<r_{isco}$
all circular orbits will fall inevitably onto the singularity, we will consider $r_{isco}$ as our
accretion radius ($r_{acc}=3r_{g}$), and all material at $r<r_{acc}$ will be considered to enter the event horizon (at $r_{g}$), and all of its properties (such as mass, angular and
linear momentum, etc...) will be transferred to the BH.

As our study is mainly focused on studying the effects of cooling and self-gravity in the
formation of instabilities at the accretion disk, we will consider a pseudo-Newtonian potential
to account for the most important general relativistic effects determining the motion of matter
near a non-rotating BH. This together with an adequate equation of state (EOS) and cooling
mechanisms will give important information about the accretion flow without solving the problem
in GR. Therefore, we will consider the BH as a $2M_{\odot}$ particle, artificially fixed at
the origin (by canceling the forces acting on it). Nevertheless, all linear and angular momentum
accreted from the gas is stored, and would be taken into account for future simulations were a
freely moving  BH will be considered. 

\begin{figure}
  \centering
 \includegraphics[width=2.35in,angle=-90]{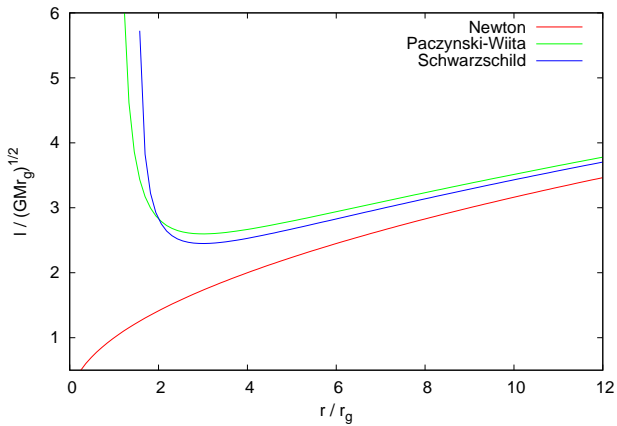}
  \label{t_nu}
  \caption{Keplerian angular momentum distribution for a test particle in a circular orbit in a Newtonian (red line), a Paczynski-Wiita (green line) and a Schwarzschild BH potential (blue line).}
\label{PWAngMom}
\end{figure}

We considered a Paczynski-Wiita (PW) potential $\Phi_{PW}$ (Paczynski \& Wiita, 1980), which
reproduces exactly, the location of the marginally stable and the innermost stable circular orbit
($r_{mb}=2r_{g}$ and $r_{isco}=3r_{g}$ respectively) for a Schwarzschild BH. 
\begin{equation}
 \Phi_{PW}=-\frac{GM_{BH}}{r-r_{g}}
\label{PWiita}
\end{equation}
This pseudo-Newtonian potential reproduces quite accurately the form of the Keplerian
angular momentum distribution $L(r)=\left(r^{3}d\phi/dr\right)^{1/2}$ obtained for a test
particle orbiting a Schwarzschild BH (See Fig~\ref{PWAngMom}). The deviation from the Schwarzschild distribution  translates into a slightly different accretion rate than the expected for
GR, given that it would have a direct effect on the angular momentum transported, which in turn,
would affect how material is transported at the disk. Nevertheless, considering a PW potential
results in a more realistic agreement with GR than considering a Newtonian potential.

When accretion of a gas particle occurs, its mass is removed from the gas and transferred to the BH.
This changes the BH properties after an integration time $dt$ on which the BH mass
will have accreted a mass $M_{acc}$ at a rate $M_{acc}/dt$. This modifies $r_{g}$ and
therefore, the accretion radius $r_{acc}$ and the PW potential at $r$. We carried out a series of tests of simple models of accretion, with and without rotation, and with and without hydrodynamical effects playing a role (by effectively reducing drastically the internal energy in the gas) to ensure that the modifications in the GADGET code accurately conserved mass, and linear and angular momentum during the accretion process.

\subsection{Thermodynamics \& Cooling Schemes}

The code GADGET-2 (Springel 2005) includes an ideal gas EOS, in terms of an entropic  function
$A(s)\equiv P/\rho^{\gamma}$. Where $\gamma=5/3$ is the adiabatic index. Therefore, if we study
the collapse of our $2.5M_{\odot}$ polytropic envelopes without the inclusion of any cooling
mechanism on the code, the collapse would be an adiabatic one. The gas would only loose energy
(internal, potential \& kinetic) when it gets accreted  by the BH. Although this scenario is not
the one we expect in a collapsar, given that there would be an important neutrino cooling at some
regions of the disk (where $T$ and $\rho$ are high enough to achieve par creation \& annihilation
or neutronization), it is important to study it to determine the importance of cooling mechanisms
in the accretion flow and the overall morphology of the accretion disk formed around the BH. 

The adiabatic collapse will give us information about the collapse and formation of
instabilities in an envelope whose cooling mechanisms are highly inefficient. On the other hand,
an isothermal collapse accounts for an undetermined cooling process which ``radiates'' energy in
order to maintain a constant temperature throughout the envelope as it collapses and orbits the
BH. In the former scenario the energy loses are immediate, and the envelope would have already
loss a great amount of energy prior disk formation. Therefore, the isothermal scheme accounts as
a collapse with a highly efficient cooling mechanism. This isothermal collapse is obtained by
setting $\gamma\simeq1$ in the EOS.

In order to explore the implications of different efficiencies in the cooling mechanism, we explored
the transition from an adiabatic collapse ($\gamma=5/3$) without cooling, to an isothermal one
($\gamma\simeq1$), in a $2.5\ M_{\odot}$ polytropic envelope with a fixed circularization radius
$r_{c}=7.5\ r_{acc}$. For $\gamma=5/3$, we adopted a simplified cooling prescription based on a
fixed cooling time $t_{cool}$ dependent of the dynamical time scale of the accretion disk
$t_{disk}$ formed around the BH.
\begin{flalign}
\qquad t_{cool}& =\beta t_{disk} \mbox{ ,}\qquad \frac{du_{i}}{dt} = u_{i}/t_{cool} 
\label{cooling}
\end{flalign}
This cooling prescription is determined by the internal energy $u_{i}$ and the efficiency parameter $\beta$, which determines how many times the gas must orbit the BH before it gets significantly cooled. If $\beta\gg1$ then the gas goes around the BH many times before losing a significant fraction of its internal energy and the infall will resemble an adiabatic one. Meanwhile, if $\beta\ll1$ then the envelope cools down before forming the disk and the cold gas would fall freely (except for rotation) onto the BH. Without compression or expansion of the envelope, the only change in the internal energy would be given by equation (\ref{cooling}) and the cooling would be exponential.

The dynamical time scale for the disk can be estimated from the initial angular velocity
$\Omega_{0}=8.33 \mbox{ s}^{-1}$ obtained for a circularization radius $r_{c}=7.5\ r_{acc}$,
such that $t_{disk}=1/\Omega_{0}=0.12\mbox{ s}$. After disk formation, the accretion disk would
have a nearly keplerian angular momentum distribution. Therefore, $\Omega_{0}$ will not represent anymore the angular velocity of the hole disk. Nevertheless, it will still give us an estimate of the time it takes the innermost material to orbit around the BH. 

\subsubsection{Neutrino Cooling Time}

In order to better understand and characterize the formation and evolution of instabilities within the disk with the presence of an effective cooling mechanism, such as neutrino emission, we explored the case
where $t_{cool}$ is close to the physical neutrino cooling time scale, $t_{\nu}$. A first approximation to
$t_{\nu}$ can be obtained from the work by Narayan, Piran \& Kumar (2001), where they
study different accretion flow scenarios onto a compact object (BH) in the context of GRB production. For a given
temperature, $T$, and density, $\rho$, the cooling rate per unit volume due to neutrinos is given
by (Narayan et al. 2001):
\begin{flalign}
  q_{\nu} \simeq5\times10^{33}T_{11}^{9}+9.0\times10^{23}\rho T_{11}^{6}\mbox{ ergs
cm}^{-3}\mbox{ s}^{-1}\label{ne_cool},
\end{flalign}
where the first term on the right hand side comes from pair annihilation and the second term from pair capture onto free nucleons (estimated for a fully photodesintegrated gas where the mass fraction of free nucleons is $X_{nuc}=1$). Both terms depend sensitively on temperature,  $T_{11}=T/10^{11}\mbox{ K}$, so high temperatures, $T\gtrsim10^{10}\mbox{ K}$, are required for this to become relevant.

\begin{figure}
\centering
 \includegraphics[width=2.35in,angle=-90]{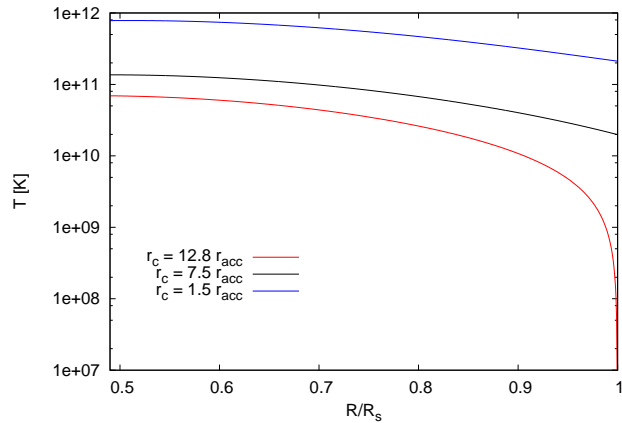}
  \label{t_nu}
  \caption{Estimated temperature $T$ at the centrifugal barrier $r=r_{c}$ as function of the
initial position of the envelope material. Each line corresponds to a different circularization
radius $r_{c}$ (or equivalently, angular velocity $\Omega_{0}$).}
\label{TvsR}
\end{figure}

We can estimate the temperature that the infalling gas would acquire upon arrival to the disk by calculating its free fall velocity $v_{ff}$ at the centrifugal barrier ($r=r_{c}$), where material would be shocked and heated. Assuming that at the shock ($r=r_{c}$) all kinetic energy is transformed
into thermal energy, we can estimate an upper limit $T_{up}$ to the temperature $T$ the
infalling gas would acquire as $T_{up}=mv_{ff}^{2}/3k$. To estimate $v_{ff}$ we will assume the Hydrogen gas has negligible pressure support and no rotation. This results in an overestimation of $v_{ff}$ since the velocity won't be completely radial when rotation is included but is a fair approximation since the rotation rate is actually quite low. So, we may plot the temperature $T_{up}$ acquired by the infalling gas at $r_{c}$ as a function of its initial radius $r_{i}$ (Figure~\ref{TvsR}). For the circularization radius used in our simulations, $r_{c}=7.5\ r_{acc}$, $T_{up}$ is $10^{10}\mbox{ K}\lesssim T \lesssim 10^{11}\mbox{ K}$. Therefore,  assuming that the infalling gas will acquire temperatures between $10^{10}\mbox{ K}\lesssim T \lesssim 10^{11}\mbox{ K}$ at the innermost
part of the disc is seen to be a good approximation.

On the other hand, the mean density of the pre-collapse polytropic envelope, $\bar{\rho}$, can be obtained from the data in Table~\ref{table1}, as $\bar{\rho}=1.7\times10^{9}\mbox{ g cm}^{-3}$. This implies that after collapsing onto the BH, with a change in spatial scale of a factor of a few to 10 at the least, material will be further compressed in some regions and can easily reach densities from 10 to 1000 times larger, depending on the local pressure and the cooling mechanism implemented. Thus densities comparable to those considered by Narayan et al. (2001), $10^{10}\mbox{ g cm}^{-3}\lesssim\rho\lesssim10^{12}\mbox{ g cm}^{-3}$ will also occur.

The internal energy of the gas will have three main contributions. Firstly, from an ideal gas composed of $\alpha$ particles and free nucleons. Secondly, a contribution from radiation, which would be effectively trapped due to the high densities reached at neutrino cooling regions (the photon mean free path is many orders of magnitude smaller than the typical length scale of the problem). And finally, a relativistic electron-positron component, which can have arbitrary degeneracy. To obtain an approximation of the expected internal energy under such conditions, we will consider that the relativistic electron-positron pairs are hot and fully non-degenerate. Thus $u_{gas}=3/2\ kT\rho/(\mu m_{p})$, and photons and pairs will contribute together a radiation energy density  $u_{rad+pair}=(11/4)\ aT^{4}$. Therefore the full internal energy of the gas (per unit volume) can be estimated by:
\begin{flalign}
  u=\frac{3}{2}\frac{kT\rho}{\mu m_{p}} + \frac{11}{4}aT^{4}\label{u_int}
\end{flalign}
From (\ref{ne_cool}) and (\ref{u_int}) and considering densities between
$\rho\simeq10^{10}-10^{12}\mbox{ g cm}^{-3}$ and temperatures of the order
$T\simeq10^{10}-10^{11}$ K, we can estimate a neutrino cooling time scale
$t_{\nu}=u/q_{\nu}$ as shown in Figure~\ref{t_nu}. This estimate shows that the neutrino cooling time scale $t_{\nu}$ will range from $\sim10^{-5}$ to a few seconds, depending mainly on the
temperature of the system. Table~\ref{table3} shows the efficiency parameter $\beta$ used in
our simulations, and the corresponding cooling time, $t_{cool}$. Our models will be
labeled after their $t_{cool}/t_{dyn}$ ratio, as shown in Table\ref{table3}.

\begin{figure}
\centering
 \includegraphics[width=2.4in,angle=-90]{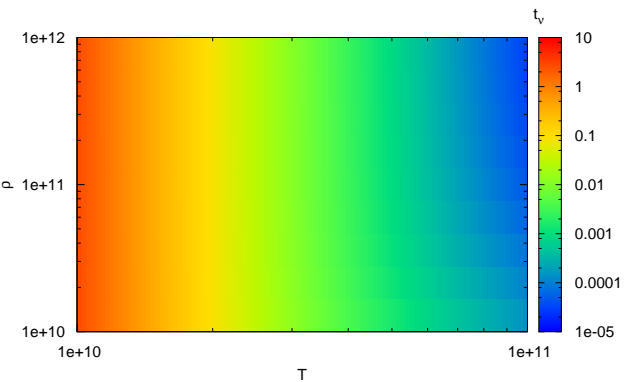}
  \caption{Neutrino cooling time scale $t_{\nu}$ estimated for an ideal gas ($\mu=1/2$) with
thermal relativistic electron-positron pairs and radiation pressure contributions.}
\label{t_nu}
\end{figure}

\begin{table}
 \begin{center}
  \begin{tabular}[!h]{l|c|c|c}
\hline
Model  & Cooling efficiency $\beta$ & $t_{cool}$& $t_{cool}/t_{dyn}$\\\hline
 $\beta$Ad  &  Adiabatic   & --    & --\\
 $\beta13.4$ & 10.27 &    1.2319 s & 13.4\\
 $\beta2.68$ & 2.054 &    0.24639 s & 2.68\\
 $\beta1.34$ & 1.027 &   0.12319 s & 1.34\\
 $\beta0.67$ & 0.513 &   0.061595 s & 0.67\\
 $\beta0.134$ & 0.1027 &  0.012319 s & 0.134\\
 $\beta0.067$ & 0.0513 &   0.0061595 s & 0.067\\
 $\beta$Iso & Isothermal & -- & --
\\\hline
\end{tabular}
\end{center}
\caption{Cooling times $t_{cool}$ and efficiency parameters $\beta$ used in our simulations. All
times range within the neutrino cooling time, $t_{\nu}$ previously estimated.}
\label{table3}
\end{table}



\section{Results \& Discussion}\label{results}

We will first discuss the general features of the accretion process, such as the BH
mass accretion rate $\dot{M}=dM/dt$, the evolution of the BH mass $M_{BH}$, and the energy loss rate $L_{c}=du/dt$, defined by the cooling times from Table\ref{table3}. These quantities will give us information on the general changes in the behavior of the system when implementing different cooling schemes. Then we will study in detail some specific models, in order to look for a relation between these quantities's behavior and the disk's specific properties, or the cooling time scale itself.

\subsection{Accretion rates and BH Mass}

Given that the polytropic envelope is intrinsically located at a distance $\sim R_{s}/2$ from
the BH at $t=0$, the accretion will not begin until the innermost material reaches the center of
the distribution. The time the envelope takes to reach the BH is affected by the gas pressure
which, in the case of being negligible, would translate into a free-fall time of the envelope
$t_{fall}\sim t_{dyn}/2$. In reality, gas pressure will make the envelope reach the BH at a slightly
earlier time, $t_{fall}\lesssim t_{dyn}/2$.

\begin{figure}
\centering
 \includegraphics[width=2.38in,angle=-90]{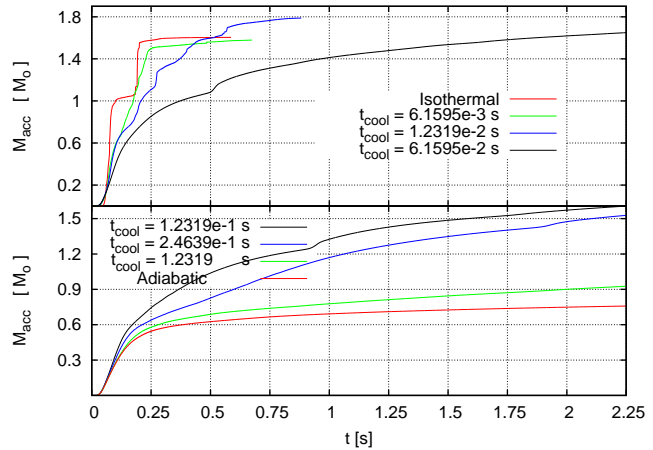}
  \caption{Accreted mass (solar masses) as function of time (seconds) for all different cooling rates 
used. In the top panel we show the envelopes with the shorter cooling time scale $t_{cool}$
which resembles  the isothermal envelope (red line) the most, while the bottom panel shows the
slowly cooled envelopes, resembling the adiabatic case (red line).}
\label{Mbh}
\end{figure}

Figure~7 shows the accreted mass as a function of time for all different models. In every case, accretion begins at a time $0.02 \mbox{ s}\lesssim t\lesssim0.05\mbox{ s}$ depending
on the cooling scheme. The more efficient the cooling, the more $t_{fall}$ and the accreted mass
increase. At the bottom panel of Figure~\ref{Mbh} we can see the slowly cooled envelopes ($\beta13.4$,
$\beta2.68$ \& $\beta1.34$) which resemble  the adiabatic one (red line) the most. All these models
show a smooth increase in the accreted mass for at least the first $\sim0.9$ seconds. Models $\beta2.68$ and $\beta1.34$ have an abrupt increase in the accreted mass after $\sim0.9$ and $\sim1.9$ s respectively. In the top panel, we can see the cooled envelopes with higher cooling efficiency ($\beta0.67$, $\beta0.134$ \& $\beta0.067$) which resemble the isothermal envelope (red line) the most. Efficiently cooled models show abrupt changes in the accreted mass and considerably higher accreted mass than the slowly cooled ones. Models with high cooling efficiency (as well as the isothermal one) are terminated at earlier times because of the computational demands imposed by an increasingly shorter time step in the final stages. Special care was taken in order to keep the gas from cooling below $u=0$, by imposing a threshold in the cooling subroutine so that the energy was lost only for gas with $T>1000$~K. 

Figure~8 shows the evolution of the accretion mass rate for all models, whose first peak at
$t\lesssim0.1$~s is due to the accretion of infalling material with low angular momentum. As
material reaches the BH, a shock is formed and propagates outwards. If no cooling is enabled, the evolution is adiabatic, which will slow down the infalling material and prevent some of it from being accreted. Meanwhile, with finite cooling, the shocked material is able to lose energy, and the shock is slowed down faster than in the adiabatic envelope. This translates into a higher accretion rate shown as a peak at $t\sim0.1$ s in Figure~8 (more distinguishable in the bottom panel). 

If the shock is strong enough to slow down the infalling material, the accretion rate will be
diminished considerably. This, together with the exhaustion of low angular momentum material,
translates into a decrease in the accretion rate, shown at times $0.1\mbox{ s}\lesssim t \lesssim
0.25\mbox{ s}$ in Figure~8. This doesn't seem to hold for models $\beta0.134$, $\beta0.067$ and $\beta$Iso
which show strong variations in $dM/dt$ after $t\simeq0.1$~s. Once the shock has passed
through the whole envelope, material will settle around the BH and form an accretion disk, whose
thickness decreases with increasing cooling efficiency, as expected. From this point on, material from the disk will be able to fall onto the BH if angular momentum transport occurs. This can be seen in the bottom panel of Figure~8 around $t=0.3$~s, when the accretion rate has decreased by nearly one order of magnitude with respect to the initial peak.

\begin{figure}
\centering
 \includegraphics[width=2.38in,angle=-90]{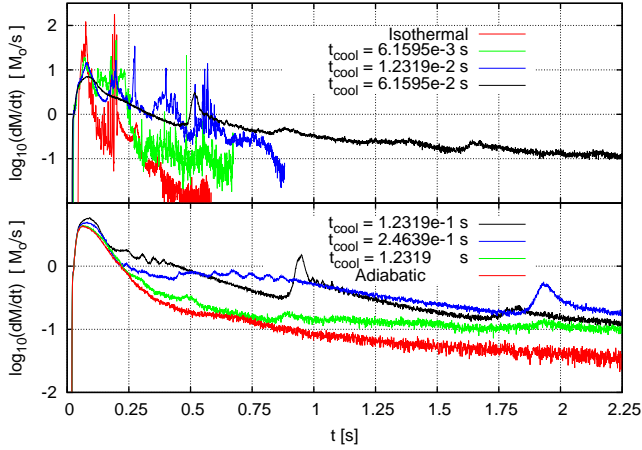}
  \caption{Logarithm of the BH mass accretion rate $\dot{M}$ in solar masses per second. In the
top panel we show the envelopes with the smaller cooling time scale $t_{cool}$, which resemble the isothermal envelope (red line) the most, while the bottom panel shows the slowly cooled
envelopes, akin to the adiabatic case (red line).}
\label{Mdot}
\end{figure}

As we can see from Figure~\ref{Mdot}, the slowly cooled models show smaller variations in the accretion
rate $\dot{M}$ than the more efficiently cooled ones. Such variations  must
be due to a change in the way material is transported within the disk. Particularly, model
$\beta0.134$ (blue line in the top panel of Figure~\ref{Mdot}) shows several peaks in $\dot{M}$ at times $t\sim0.2, 0.3, \mbox{ and }0.4$~s. Some quasi-periodic pattern may be present in the accretion rate of model $\beta0.134$, and it will be studied carefully below.

\subsection{Cooling Efficiency and Heat Losses}

Models implemented with our cooling prescription will be loosing energy each time step at a rate
$du_{i}/dt = u_{i}/t_{cool}$. This cooling rate  depends on the internal energy $u_{i}$, at
the position $(r,\phi,z)$. Therefore, energy losses won't be uniform throughout time and/or
space. Considering the contribution from each SPH particle, we can obtain the
total energy loss rate $L_{c}=\sum_{j}(du_{i}/dt)_{j}$ at a given time $t$.

\begin{figure}
\centering
 \includegraphics[width=2.38in,angle=-90]{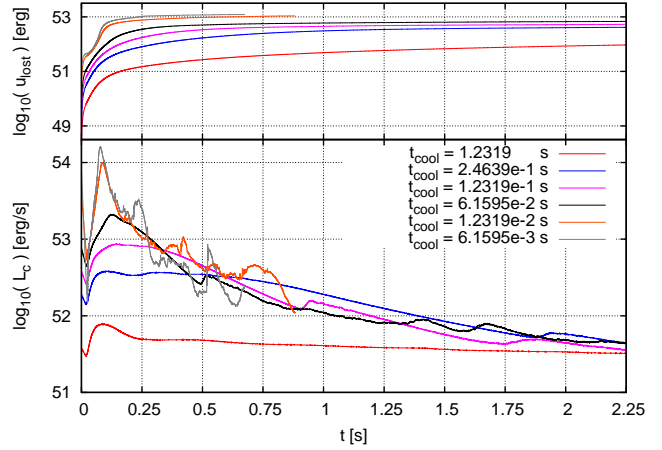}
  \caption{Logarithm of the energy loss rate $L_{c}$ (bottom panel) in erg/s for models
$\beta13.4$ (red line), $\beta2.68$ (blue line), $\beta1.34$ (pink line), $\beta0.67$ (black line), $\beta0.134$ (orange line) and $\beta0.067$ (gray line). In the top panel we show the integrated energy lost by all SPH particles, $u_{lost}$, as a function of time.}
\label{Lum}
\end{figure}

Energy loss rates $L_{c}$ (i.e., neutrino luminosities) for different cooling efficiencies are showed in the bottom panel of Figure~\ref{Lum}. The maximum $L_{c}$ is obtained for the model with higher cooling efficiency, reaching up to $L_{c}\sim10^{54}\mbox{ erg s}^{-1} = 1000\mbox{ foe s}^{-1}$. The total energy lost $u_{lost}=\int L_{c} \ dt$ is shown in the top panel of Figure~\ref{Lum}, with a maximum value $u_{lost}\sim10^{53}$ erg that is also reached by the most efficiently cooled model, in only $t\lesssim0.7\mbox{ s}$. 

As the cooling efficiency increases, the changes in the energy loss rates $L_{c}$ become more
drastic. By comparing Figures~\ref{Mdot} and \ref{Lum}, we see that such abrupt changes coincide in time with intense increases in the accretion rates. Given that a rapid increase in the energy loss rate $L_{c}$ is only possible if there is an increase in the internal energy ($du/dt = u/t_{cool}$), some mechanism must be causing the increase in the local internal energy $u$. As these intense variations on both $\dot{M}$ and $L_{c}$ appear as the cooling efficiency is improved, it seems logical to assume that the formation of structure within the accretion disk is triggering these variations. Therefore, we study  in detail the formation of structures, such as spiral arms, in order to look for a coincidence of structure formation events with the intense increases in both the accretion rate and the energy loss rate. 

Based on the obtained energy loss rates $L_{c}$ from our different models, and the
assumption that the neutrino cooling would have a cooling time scale $t_{\nu}$ ranging within
the cooling times $t_{cool}$ explored here, we can expect to obtain neutrino luminosities
$L_{\nu}$ ranging between $1\lesssim L_{c}\lesssim2000\mbox{ foe s}^{-1}$ (depending on the local
temperature and density conditions) when a realistic neutrino prescription is used. An efficient
neutrino cooling will be restricted to regions with really high temperatures ($T\gtrsim10^{11}$
K) and densities ($\rho\gtrsim10^{10}$), and this cooling mechanism won't be efficient during the
initial collapse of the envelope, when the bulk of the gas has not yet reached the BH. This may
affect the energy loss rates significantly at earlier times. Moreover, here we do not consider the possibility of energy deposition from the cooling mechanism itself, which will be important for neutrino cooling at regions of high density ($\rho\gtrsim10^{12}\mbox{ g cm}^{-3}$) when the opacity due to neutrons and $\alpha$ particles becomes important (Shapiro \& Teukolsky 1983). This may diminish the neutrino luminosity and could also prevent the material from cooling down so efficiently.

\subsection{Characteristic Time Scales in the Disk}

\begin{figure}
\centering
  \includegraphics[height=3.7in,angle=-90]{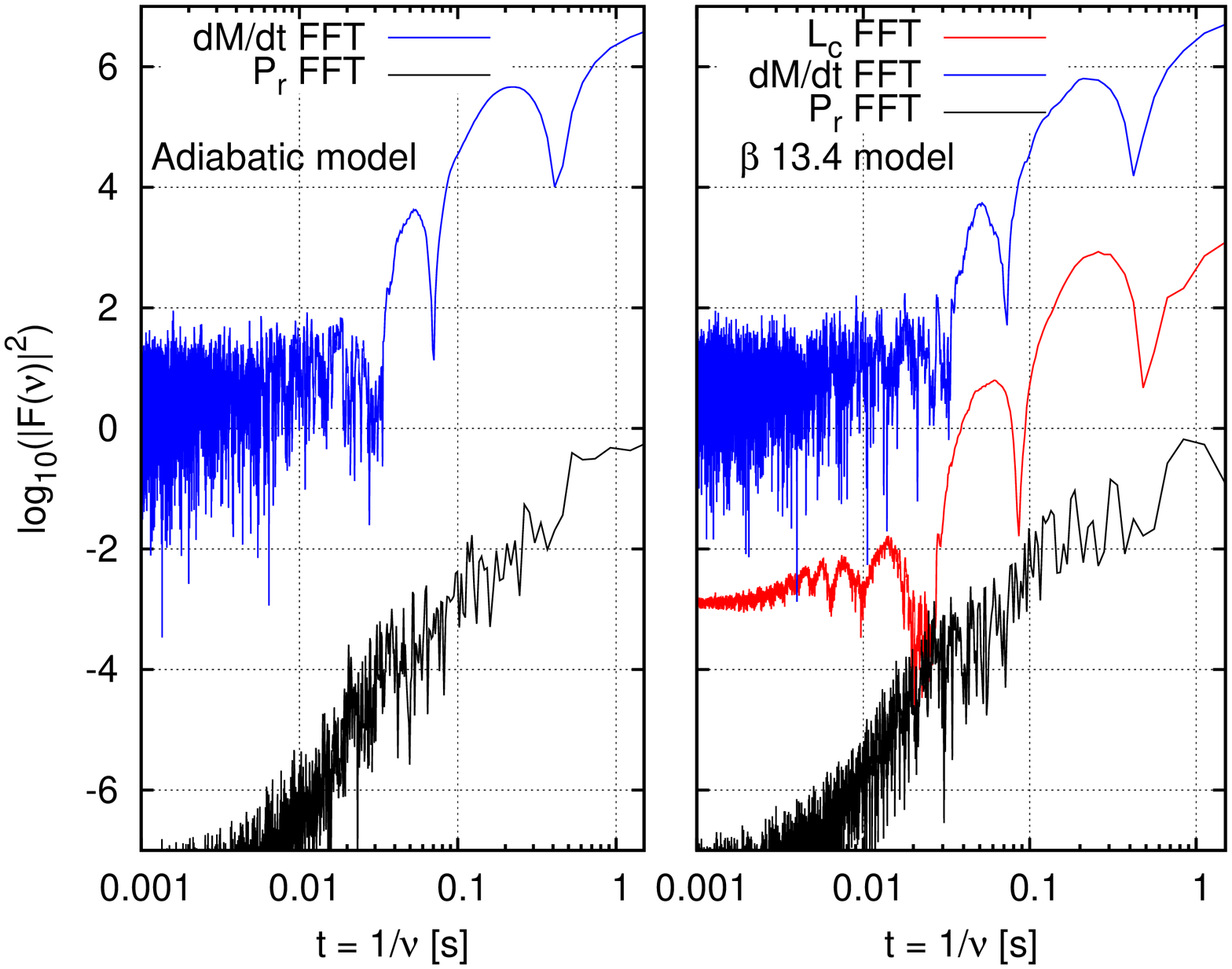}
  \includegraphics[height=3.7in,angle=-90]{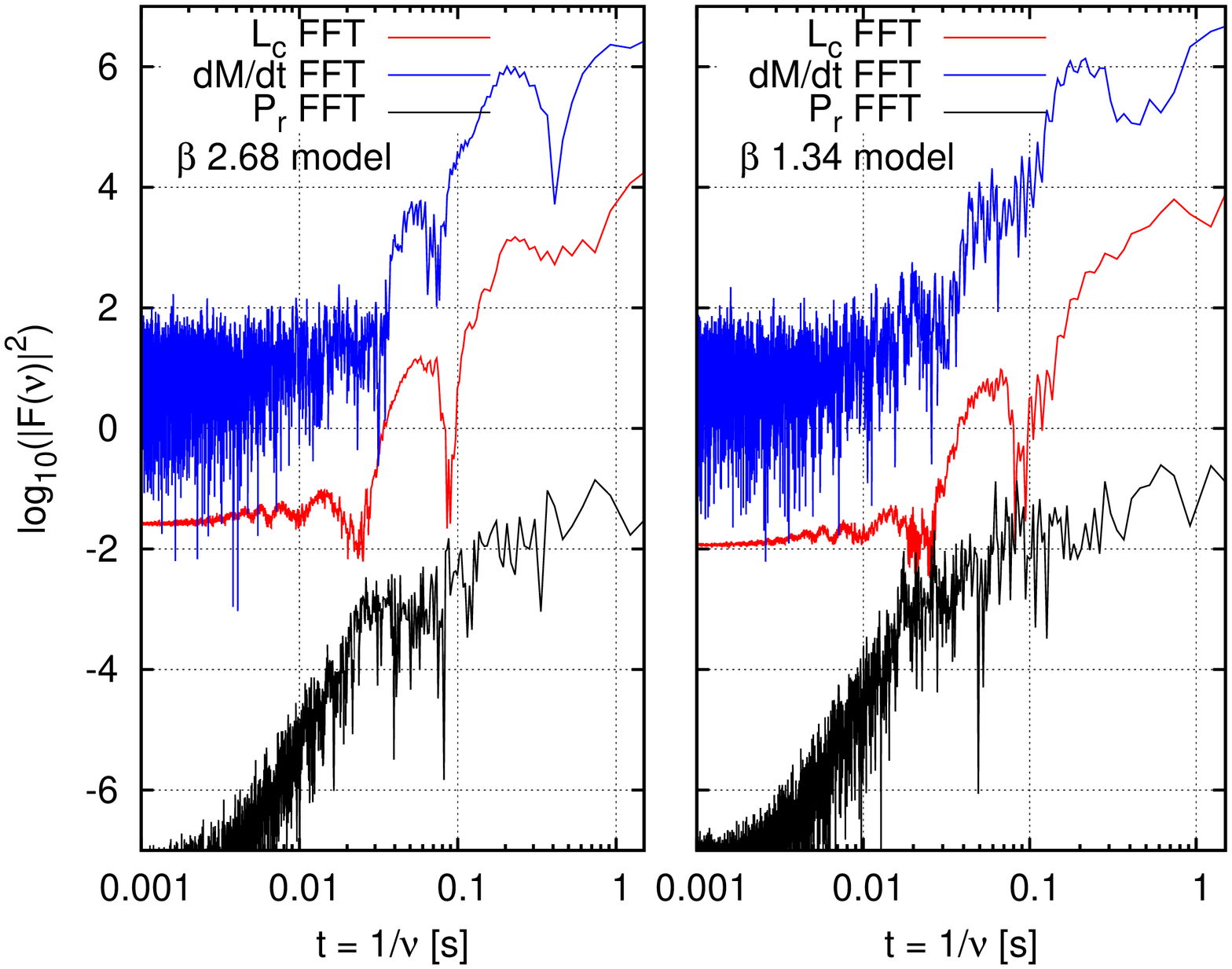}
  \caption{Logarithm of the renormalized Fourier's transform power $|F(\nu)|^{2}$ of the mass accretion
rate $\dot{M}$ (blue line), the energy loss rate $L_{c}$ (red line) and the cylindrical radial component of momentum $P_{r}$ (black line) versus inverse frequency $1/\nu$ (time), for the adiabatic case, and models $\beta13.4$, $\beta2.68$ and $\beta1.34$ (we only show the transform of $L_{c}$ for cooled models). The spectrum of the mass accretion rate $\dot{M}$ shows power at intrinsic time scales of $t\simeq0.2 \mbox{ and }0.05$~s that also appear in the spectrum of $L_{c}$. Increasing the cooling efficiency alters the position and intensity of the peaks time scales and in model $\beta1.34$ an additional peak is present at  $t\simeq0.02$~s. }
\label{Fou_1}
\end{figure}

\begin{figure}
\centering
  \includegraphics[height=3.7in,angle=-90]{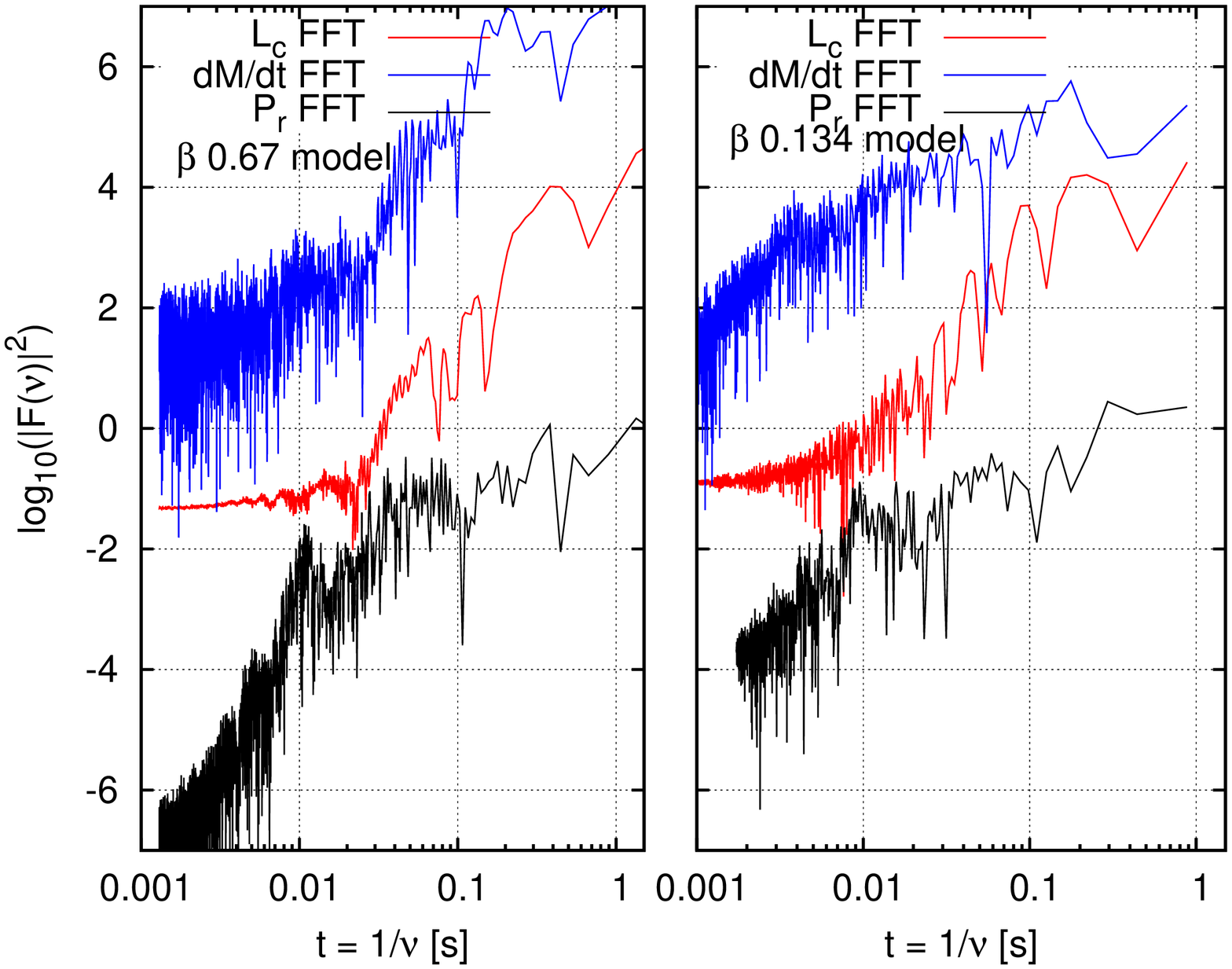}
    \includegraphics[height=3.7in,angle=-90]{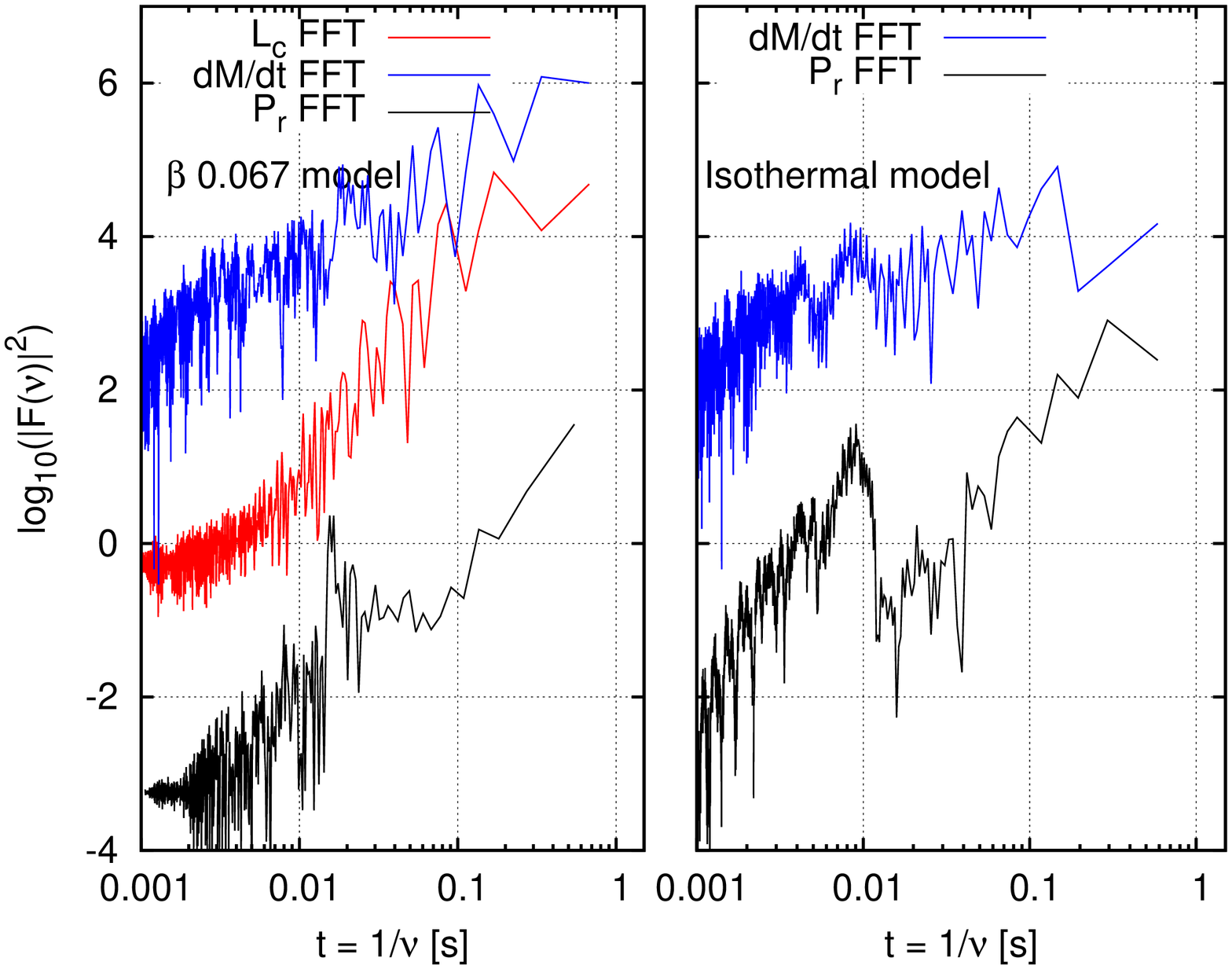}
  \caption{Logarithm of the renormalized Fourier's transform power $|F(\nu)|^{2}$ of the mass accretion
rate $\dot{M}$ (blue line), the energy loss rate $L_{c}$ (red line) and the cylindrical radial component of momentum $P_{r}$ (black line) versus $1/\nu$ (time), for models $\beta0.67$, $\beta0.134$, $\beta0.067$ and the isothermal case. More efficient cooling seems to increase the number of intense peaks appearing also at shorter time scales (higher frequencies).}
\label{Fou_2}
\end{figure}

In order to look for characteristic time scales for the gas in the accretion disk, we performed  Fourier transforms of the time series for  $\dot{M}$, $L_{c}$ and the cylindrical radial component of the momentum $P_{r}$. Any characteristic time scale occurring in them will appear as a peak in the Fourier transform amplitude $|F(\nu)|^{2}$. Figures~\ref{Fou_1} and \ref{Fou_2} show the Fourier transforms for $L_{c}$ (red line), $\dot{M}$ (blue line) and $P_{r}$ (black line) for all models (except $L_{c}$ for the adiabatic and isothermal models) as a function of inverse frequency  $t = 1/\nu$ (in seconds).

Figure~\ref{Fou_1} shows that there is a transition in the peaks in $|F(\nu)|^{2}$ for $\dot{M}$ appearing at $t\simeq0.2 \mbox{ and }0.05$~s when the cooling efficiency $\beta$ increases. These two peaks in $\dot{M}$ also appear in $L_{c}$ for models $\beta13.4$ and $\beta2.68$, but the peak at $t\simeq0.05$~s is not that clear for model $\beta1.34$. The transform for $L_{c}$ also shows less intense peaks at time scales shorter than 0.02~s and longer than 0.001~s. 

There are more characteristic time scales appearing on the cylindrical radial component of the momentum $P_{r}$, and they become clearer on models with more efficient cooling than $\beta2.68$, with characteristic time scales ranging from $t\simeq0.02$ to $0.2$~s. This range  partly coincides with that associated to the epicyclic frequency due to the BH, which varies  approximately from 0.001 to 0.1 seconds along the disk. There also seems to be an intense radial oscillation in the disk, seen in the transform for model $\beta0.67$,with period $T\simeq 0.08$~s, as well as another oscillation of smaller intensity with period $T\simeq0.01$~s. Models with higher cooling efficiency also show the presence of a characteristic period in the range $0.007\lesssim t\lesssim 0.015$~s.

Comparing Figures~\ref{Fou_2} and \ref{Fou_1} we can see that increasing the cooling efficiency increases the number of peaks appearing at higher frequencies (smaller characteristic time scales) in the Fourier transforms. This coincides with the increase of variations in $\dot{M}$ and $L_{c}$ seen when increasing the cooling efficiency. We note that only the more efficiently cooled models show significant variations on time scales shorter than 10 milliseconds, and as stated in the previous sections, these variations seem to be produced by the formation of instabilities in the disk.

\subsection{Morphological Features}

If the formation of patterns in the disk is responsible for the rapid variations in both $\dot{M}$ and $L_{c}$, we must start by looking for instabilities that may gave rise to the formation of such structures. This can be done by looking for regions where local gravitational instability arises because of the high density and/or low pressure (internal energy). This may be  observable in density and/or internal energy maps in the disk plane. Moreover, the appearance of any asymmetric structures in the disk could alter the angular momentum distribution and transport. Spiral arms and high-density gas clumps can alter the gravitational interaction between the disk and the black hole and lead to the transport of angular momentum across the disk, independently of the presence of any viscous mechanism (such as the magneto-rotational instability, MRI).  

In order to characterize nonaxisymmetric instabilities, we performed a one-dimensional Fourier
transform of the azimuthal distribution of mass
$\Phi_{M}=\int\left[\int\rho(\phi,r,z)dz\right]r\ dr$, as in the work of Zurek \& Benz (1986),
defining the amplitude of the $m$-th mode by:
\begin{equation}
 C_{m}=\frac{1}{2\pi}\int_{0}^{2\pi}e^{im\phi}\Phi_{M}d\phi.
\end{equation}
The power in each mode,  $|C_{m}|^{2}$, will give us information about the presence of over-dense structures with $2\pi/m$ azimuthal symmetry in  the disk. Therefore, the relative power
$|c_{m}|^{2}=|C_{m}|^{2}/|C_{0}|^{2}$ will give us information on the intensity of $m$ spiral arms compared to the integrated disk mass $C_{0}$, at time $t$. By plotting the evolution in time of such relative powers $|c_{m}|^{2}$ we can study the evolution of the disk and look for the formation and disruption of spiral arms forming at the disk.

Such spiral structures should also be observable in density maps at the $z=0$ plane or, more appropriately, in the surface density, $\Sigma(r,\phi)=\int\rho(r,\phi,z)dz$, maps as over-dense regions.
Nevertheless, in order to obtain quantitative information about the formation of structure in our
simulations, we have found it more useful to study the evolution of the Toomre parameter in the disk (Toomre 1964), which provides insight on the conditions under which the accretion disk around the BH
becomes unstable. The Toomre parameter is given by:
\begin{equation}
 Q_{T} = \frac{\kappa c_{s}}{\pi G\Sigma}
\end{equation}
where, $\kappa=(\partial W/\partial r)^{1/2}$ is the epicyclic frequency of motion for material in the disk, obtained from first
order perturbations and  subject to the effective potential $W(r)=\phi(r) +
l^{2}/2r^{2}$, $\Sigma(r,\phi)$ is the surface density of the disk, and $c_{s}$ is the
local sound speed. By evaluating the Toomre parameter $Q_{T}$ throughout the disk, we should be able to notice the formation of any spiral structures, as a region with $Q_{T}$ lower than the average value, given the higher surface density caused by its collapse. Considering that we are only using $Q_{T}$ as a parameter to visualize collapsing regions with significantly higher density and/or lower temperature, we will consider an approximation to the epicyclic frequency which neglects the contribution of the disk to the gravitational potential. The epicyclic frequency of a gas particle orbiting a BH with a PW potential is given by:
\begin{equation}
 \kappa=\left(\frac{\partial W}{\partial r}\right)^{1/2}
       =\left(\frac{GM_{BH}(r-3r_{g})}{r(r-r_{g})^{3}}\right)^{1/2}.
\end{equation}
 
This approximation will translate in values for $\kappa$ that are 1.5 to 3.5 times smaller than those obtained directly from the simulation data, because of the self-gravity of the disk, but it will be significantly simpler to evaluate them uniformly thus. In the next section section we will study the morphology of some models individually, in order to obtain information on the importance of structure formation for the variability in the mass accretion and energy loss rates. We will focus on the models with $t_{cool}\lesssim0.2$~s, which show the most significant variability.

$\beta1.34$ \textbf{Model}

\begin{figure}
 \begin{center}
  \centering
  \includegraphics[width=2.30in,angle=-90]{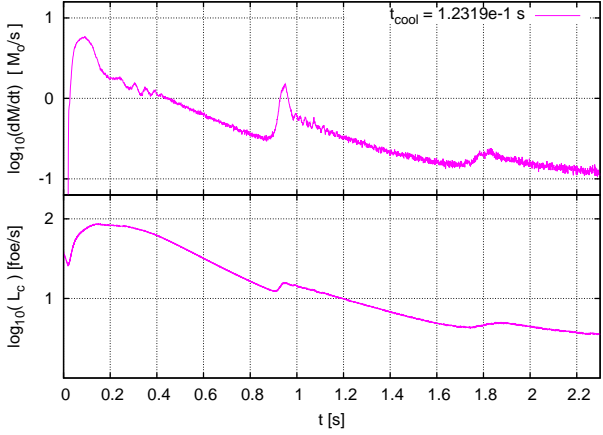}
  \caption{Mass accretion and energy loss rates for model $\beta1.34$ (top and bottom panel respectively). There is an intense increase in both $L_{c}$ and $\dot{M}$ at $t\simeq 0.9\mbox{ s }$. }
  \label{fig:MLb134}
 \end{center}
\end{figure}

\begin{figure}
\begin{center}
 \includegraphics[height=3.0in,angle=-90]{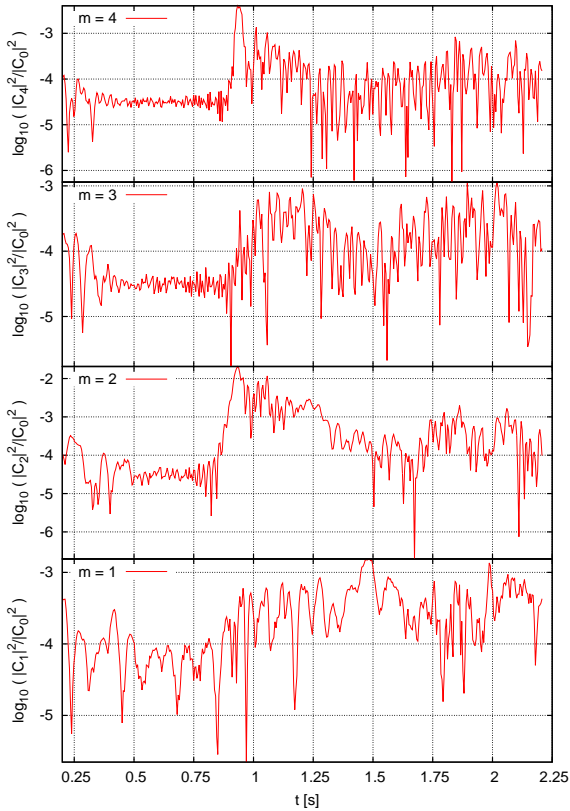}
  \caption{Relative power $|c_{m}|^{2}=|C_{m}|^{2}/|C_{0}|^{2}$ ($m=1,2,3,4$)
for the azimuthal mass distribution $\Phi_{M}$ Fourier transform in model $\beta1.34$. At $t\gtrsim0.75$ s modes $m=2, $ begin to rise, and peak at $\sim0.9$ s.}
\label{fig:Fourierb134}
\end{center}
\end{figure}

As with model $\beta2.68$, this model, with $t_{cool}=0.12319$ s, shows an abrupt change in both the
accretion rate and the energy loss rate, but at an earlier time, $t\simeq0.9$ s (Figure~\ref{fig:MLb134}). Therefore, it also suggests there should be a structure formation event producing such variations. In Figure~\ref{fig:Fourierb134} we see the evolution in time of the relative power $|c_{m}|^{2}$ for the modes $m=1,2,3\mbox{ and }4$, and there is an intense structure formation event starting at $t\simeq0.8$ s, shown in every Fourier mode in Figure~\ref{fig:Fourierb134}. Nevertheless, the $m=2$ mode seems to be the most intense of all, and therefore, we expect to observe two spiral arms on the Toomre parameter plots at $t\simeq0.9$ s when it reaches its maximum.

\begin{figure}
\begin{center}
 \includegraphics[width=3.48in]{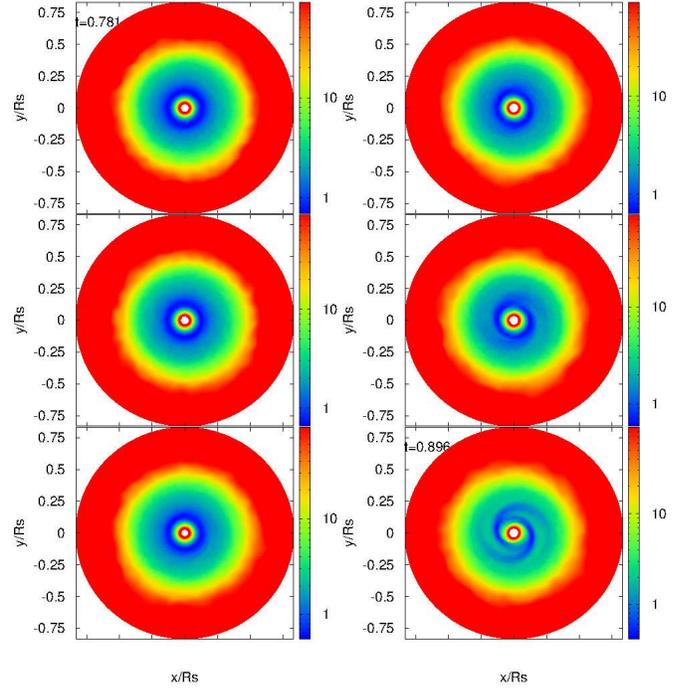}
  \caption{Evolution of the Toomre parameter $Q$ for model $\beta1.34$ at times $0.78\mbox{ s}\lesssim t\lesssim0.89\mbox{ s}$. The initial and final times are indicated on the top left panel and the right bottom panel respectively, and the evolution of time goes from top to bottom and from left to right. The last panels show the formation of two spiral arms at $t\simeq0.89$ s.}
  \label{fig:Toomre1b134}
\end{center}
\end{figure}

\begin{figure}
\begin{center}
 \includegraphics[width=3.48in]{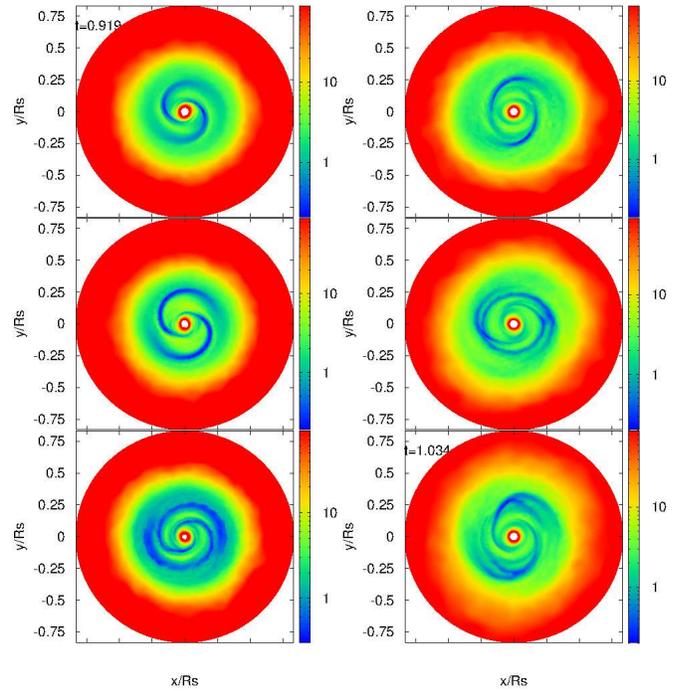}
  \caption{Evolution of the Toomre parameter $Q$ for model $\beta1.34$ at times $0.91\mbox{ s}\lesssim t\lesssim1.03\mbox{ s}$. The spiral arms remain intense for $\sim0.1$ s and break into several structures on the last panel.}
\label{fig:Toomre2b134}
\end{center}
\end{figure}

\begin{figure}
\begin{center}
 \includegraphics[width=3.48in]{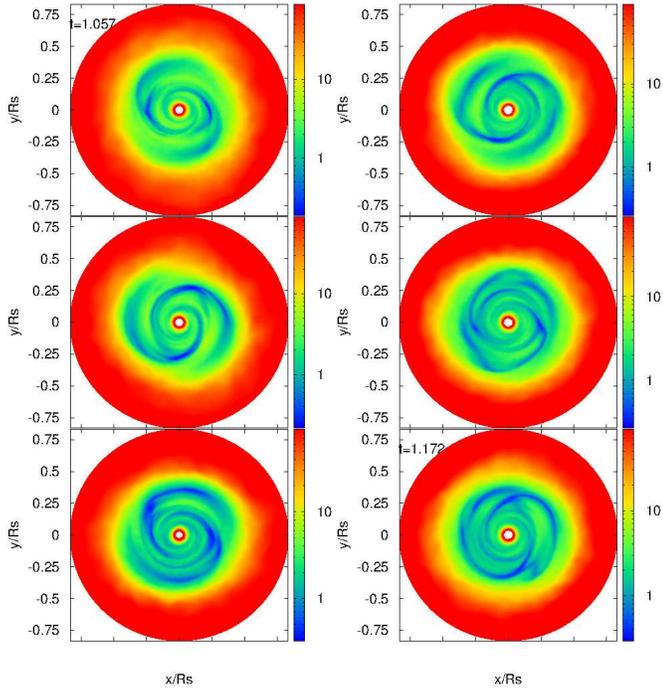}
  \caption{Evolution of the Toomre parameter $Q$ for model $\beta1.34$ at times $1.05\mbox{ s}\lesssim t\lesssim1.17\mbox{ s}$. The two spiral arms are now barely noticeable.}
\label{fig:Toomre3b134}
\end{center}
\end{figure}

\begin{figure}
\begin{center}
 \includegraphics[width=3.48in]{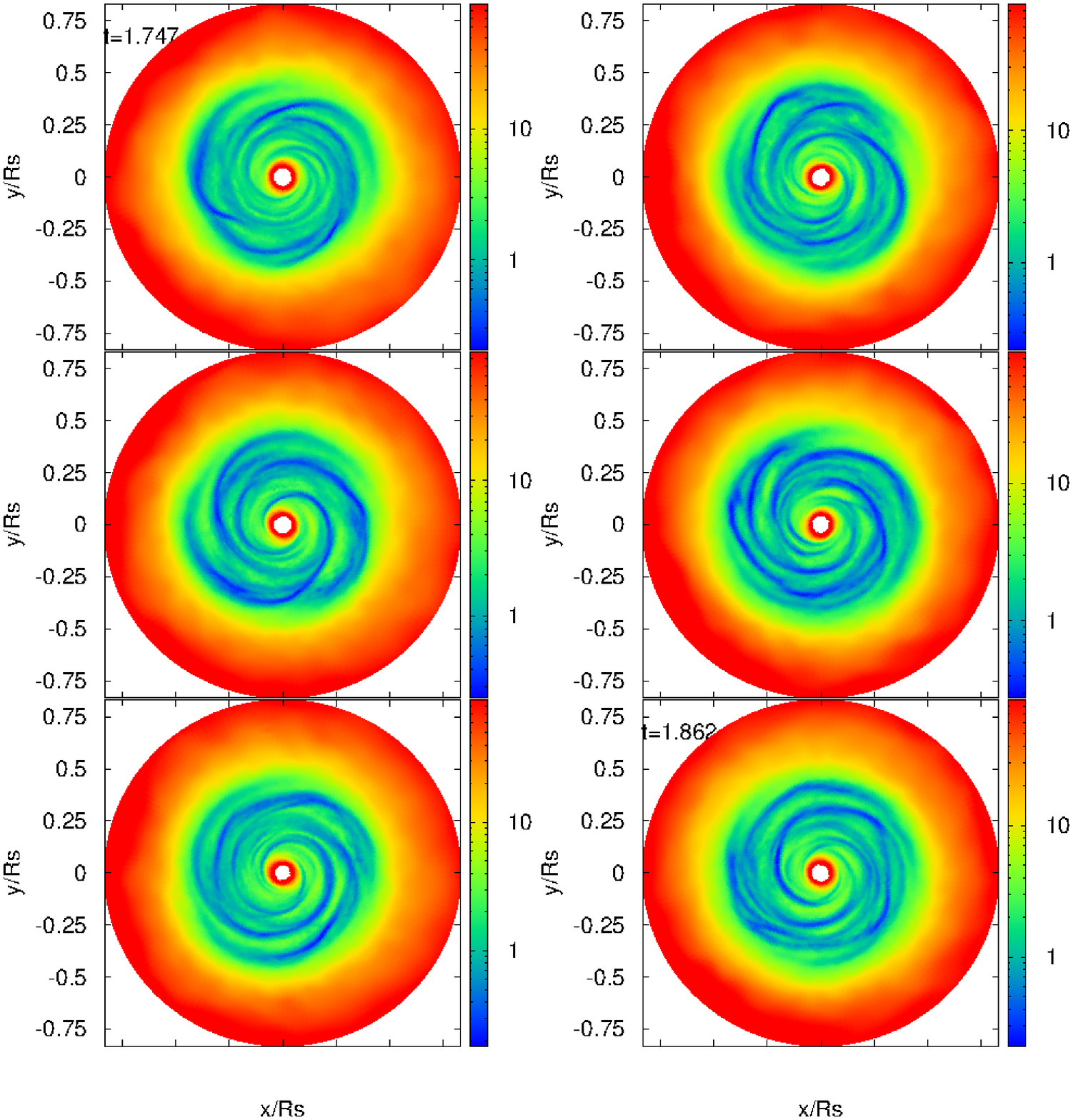}
  \caption{Evolution of the Toomre parameter $Q$ for model $\beta1.34$ at times $1.74\mbox{ s}\lesssim t\lesssim1.86\mbox{ s}$. There are several spiral structures at the disk which are shown as intense variations in modes $m=2,3$.}
\label{fig:Toomre4b134}
\end{center}
\end{figure}

Figures~\ref{fig:Toomre1b134}, \ref{fig:Toomre2b134} and \ref{fig:Toomre3b134} show the evolution in time of the Toomre parameter $Q$ for times $0.78\mbox{ s}\lesssim t\lesssim1.2\mbox{ s}$ where the mode $m=2$ reaches its maximum. Last panel from Figure~\ref{fig:Toomre1b134} shows the appearance of two spiral arms at $t\simeq0.9$ s. These spiral arms are disrupted at $t\simeq1$ s, but they are followed by the appearance of another two spiral arms on the following Toomre parameter maps. This is also visible on the Fourier mode $m=2$ showing strong variations at such times. Figure~\ref{fig:Toomre4b134} shows the Toomre parameter $Q$ at times $1.74\mbox{ s}\lesssim t\lesssim1.86\mbox{ s}$ s, where an increase on both $\dot{M}$ and $L_{c}$ can be seen in Figure~\ref{fig:MLb134}. This increase is also shown in the Fourier modes in Figure~\ref{fig:Fourierb134} as strong variations in modes $m=2,3$.

There seems to be a relationship between the time of the first structure formation event and the cooling time, but the formation of structures should also depend on the amount of mass contained in the disk. As increasing the cooling efficiency allows more material to be accreted during the initial collapse, the disk's mass can be considerably reduced and therefore, the first structure formation event could be delayed due to the fact that the disk does not have enough material to become unstable until enough cooling has taken place. 

$\beta0.67$ \textbf{Model}

\begin{figure}
 \begin{center}
  \includegraphics[width=2.30in,angle=-90]{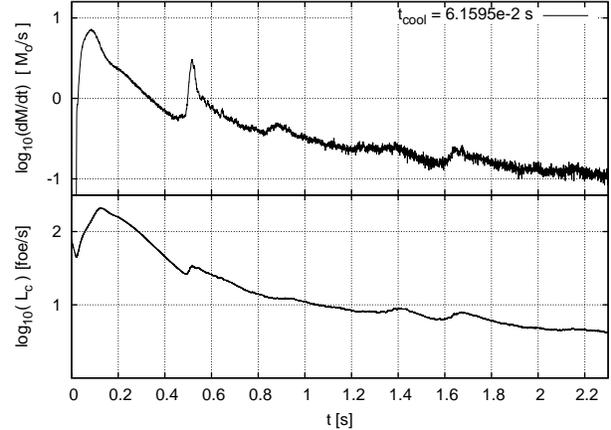}
  \caption{Mass accretion and energy loss rates for model $\beta0.67$ (top and bottom panel respectively). There are intense variations in both $L_{c}$ and $\dot{M}$ at times $t\simeq 0.5\mbox{ and } 1.6$ s.}
  \label{fig:MLb067}
 \end{center}
\end{figure}

\begin{figure}
\begin{center}
 \includegraphics[height=3.0in,angle=-90]{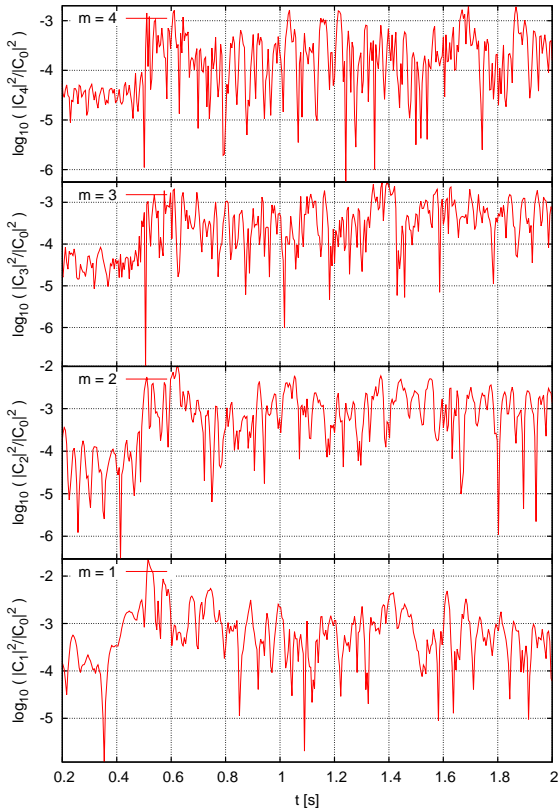}
  \caption{Relative power $|c_{m}|^{2}=|C_{m}|^{2}/|C_{0}|^{2}$ ($m=1,2,3,4$) for the azimuthal mass distribution $\Phi_{M}$ Fourier transform of model $\beta0.67$. All modes show an important increase at $t\simeq0.5$ s. Modes $m=2, 1$  reach the highest values at times $0.55\mbox{ s}\lesssim t\lesssim0.7\mbox{ s}$.}
  \label{fig:Fourierb067}
\end{center}
\end{figure}

As can be seen in Figure~\ref{fig:MLb067}, this model, with $t_{cool}=0.0616$ s shows at least two events where both $\dot{M}$ and $L_{c}$ have intense and rapid increases. Comparing with Figure~\ref{fig:Fourierb067} where we plot the evolution in time of the relative power $|c_{m}|^{2}$ for the modes $m=1,2,3\mbox{ and }4$, we see that the first event can be related to an increase in all Fourier modes beginning at $t\simeq0.45$ s. At that time, all Fourier modes seem to be have comparable power, but modes $m=1$ and 2, peaking at $t\simeq0.5$ s reach the highest values and have the longest duration. We therefore expect to see a prominent presence of one or two spiral arms, that will remain strong for a time $\lesssim0.05$ s. Comparing models $\beta1.34$ and $\beta0.67$ we see that changing the cooling efficiency, alters the width of the peaks in the Fourier power, which seems to be smaller (i.e., more sharply defined in time) with increasing cooling efficiency.

\begin{figure}
\begin{center}
 \includegraphics[width=3.48in]{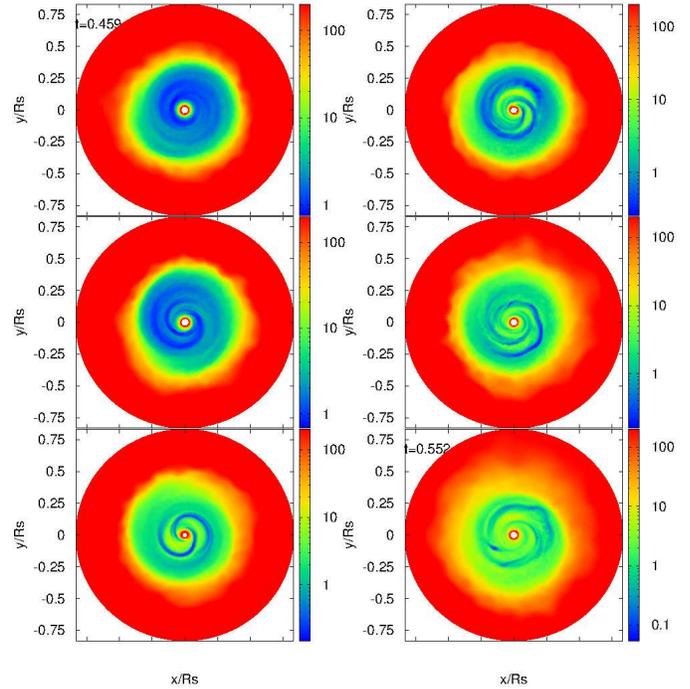}
  \caption{Evolution of the Toomre parameter $Q$ for model $\beta0.67$ at times $0.45\mbox{ s}\lesssim t\lesssim0.55\mbox{ s}$. Formation of structure is triggered at early times but does not survive for long. Transient spiral arms corresponding to Fourier modes $m=1,2$ rapidly form and disappearl.}
  \label{fig:Toomre1b067}
\end{center}
\end{figure}

\begin{figure}
\begin{center}
 \includegraphics[width=3.48in]{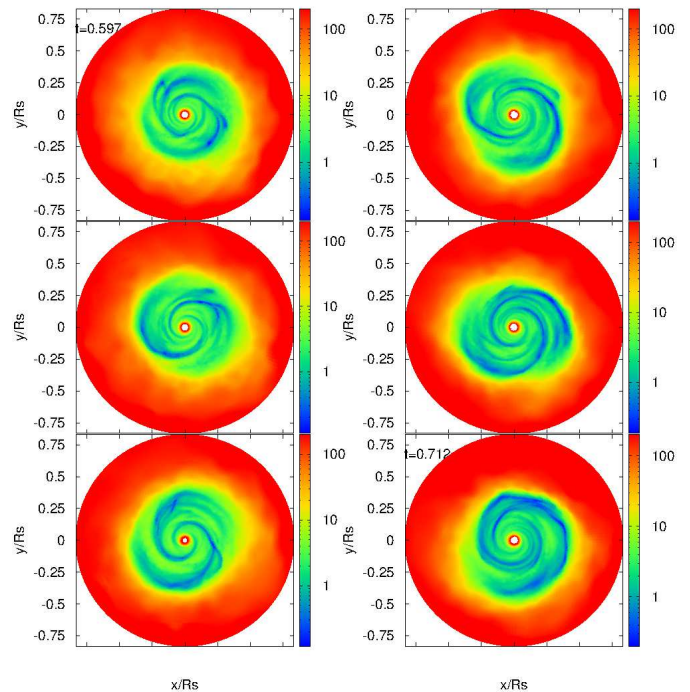}
  \caption{Evolution of the Toomre parameter $Q$ for model $\beta0.67$ at times $0.59\mbox{ s}\lesssim t\lesssim0.712\mbox{ s}$. Spiral arms break down into smaller and more collapsed structures ($Q\lesssim 0.1$) such as clumps.}
  \label{fig:Toomre2b067}
\end{center}
\end{figure}

\begin{figure}
\begin{center}
 \includegraphics[width=3.48in]{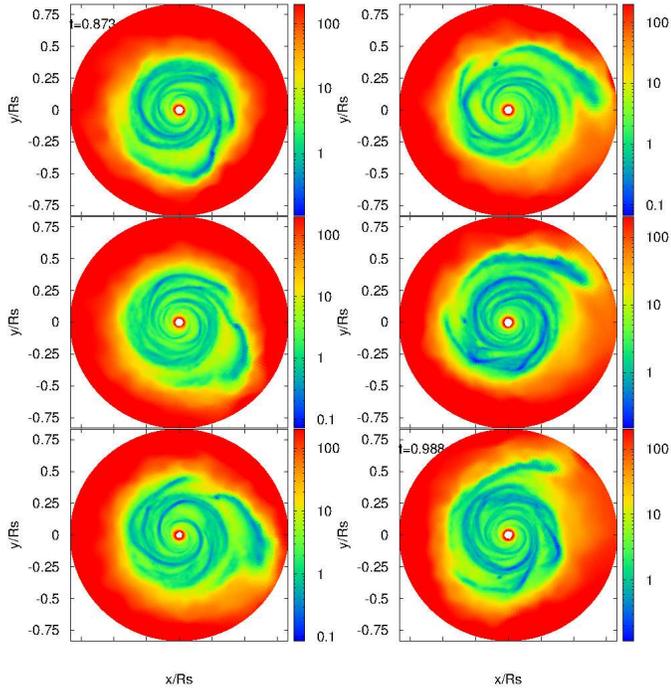}
  \caption{Evolution of the Toomre parameter $Q$ for model $\beta0.67$ at times $0.87\mbox{ s}\lesssim t\lesssim 0.98\mbox{ s}$. There are noticeable spiral arms and clumps with $Q\lesssim0.01$.}
  \label{fig:Toomre3b067}
\end{center}
\end{figure}

Figures~\ref{fig:Toomre1b067}, \ref{fig:Toomre2b067} and \ref{fig:Toomre3b067} show the evolution in time of the Toomre parameter $Q$ in the range $0.45\mbox{ s}\lesssim t\lesssim0.98\mbox{ s}$. We see in Figure~\ref{fig:Toomre1b067} that at about $t=0.5$ s one and two spiral arms are formed and disrupted in less than $0.1$ s. Remarkably, these spiral structures then break up into small and dense gas clumps, noticeable in Figure~\ref{fig:Toomre3b067}. These then interact with the rest of the spiral pattern, disrupting it and further breaking it up.

\begin{figure}
\begin{center}
 \includegraphics[width=3.48in]{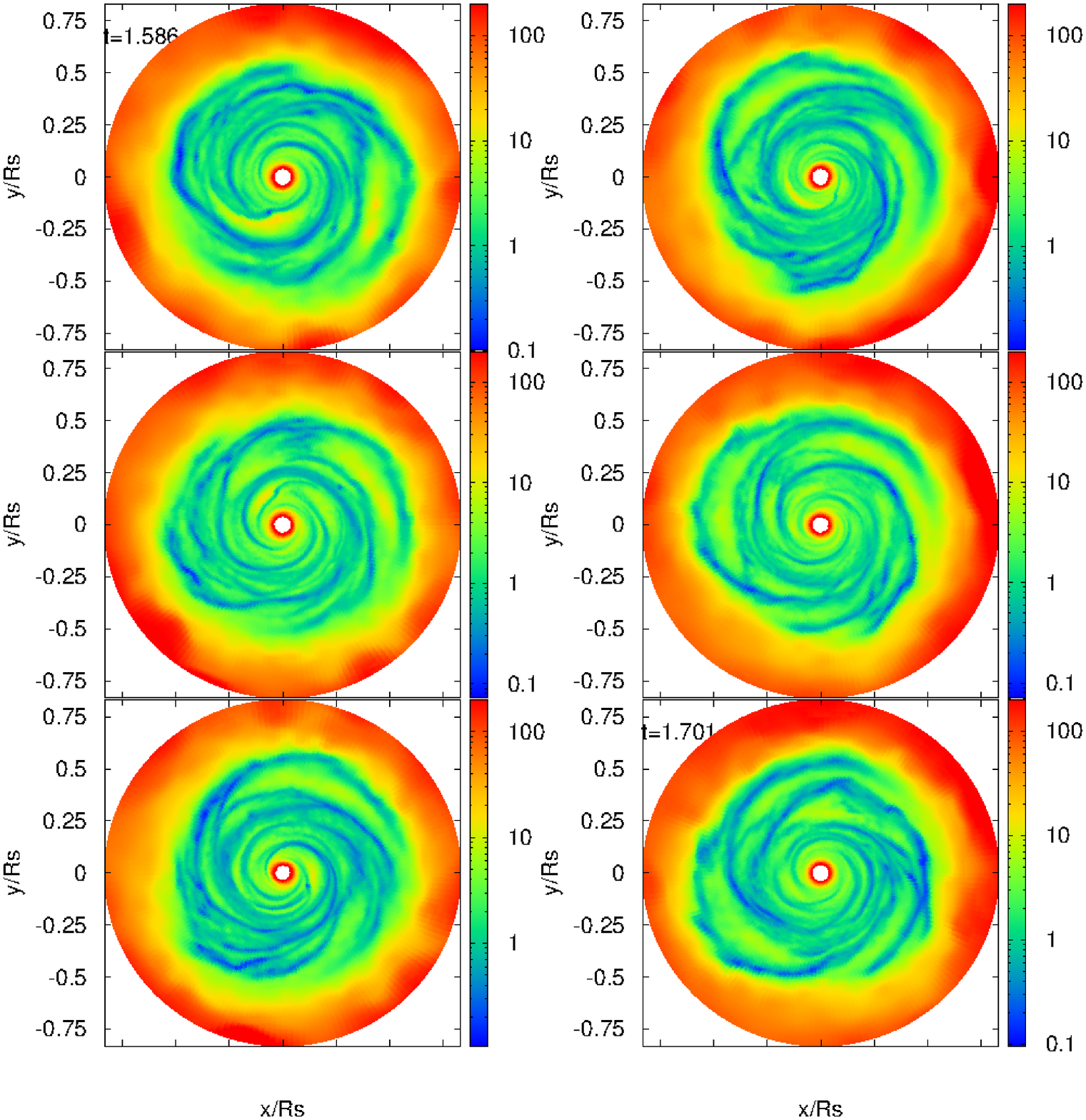}
  \caption{Evolution of the Toomre parameter $Q$ for model $\beta0.67$ at times $1.58\mbox{ s}\lesssim t\lesssim 1.70\mbox{ s}$. At this time we observe an increase in both $L_{c}$ and $\dot{M}$ and there is intense spiral structure and some gas clumps with $Q\lesssim0.1$.}
  \label{fig:Toomre4b067}
\end{center}
\end{figure}

As can be seen in Figure~\ref{fig:MLb067}, there is another important variation in $\dot{M}$ and $L_{c}$ at a times $1.6\mbox{ s}\lesssim t\lesssim1.8\mbox{ s}$, when all modes reach at some point values $|c_{m}|^{2}\gtrsim10^{-3}$. In this event all peaks are extremely narrow, because the pattern is  rapidly broken up, possibly due to  presence of gas clumps formed at earlier times. Figure~\ref{fig:Toomre4b067} shows the evolution of $Q$ for $1.58\mbox{ s}\lesssim t\lesssim1.70\mbox{ s}$, also showing clumps and spiral patterns in the disk.

$\beta0.134$ \textbf{Model}

\begin{figure}
 \begin{center}
  \includegraphics[width=2.30in,angle=-90]{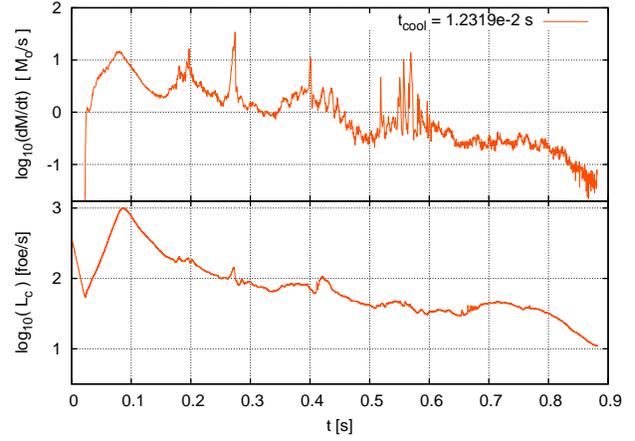}
  \caption{Mass accretion and energy loss rates for model $\beta0.134$ (top and bottom panels respectively). There are several intense changes in both $L_{c}$ and $\dot{M}$ starting at $t\simeq0.15$ s. We note that some of the intense variations in $\dot{M}$ are barely visible in $L_{c}$.}
  \label{fig:MLb0134}
 \end{center}
\end{figure}

\begin{figure}
\begin{center}
 \includegraphics[width=3.0in]{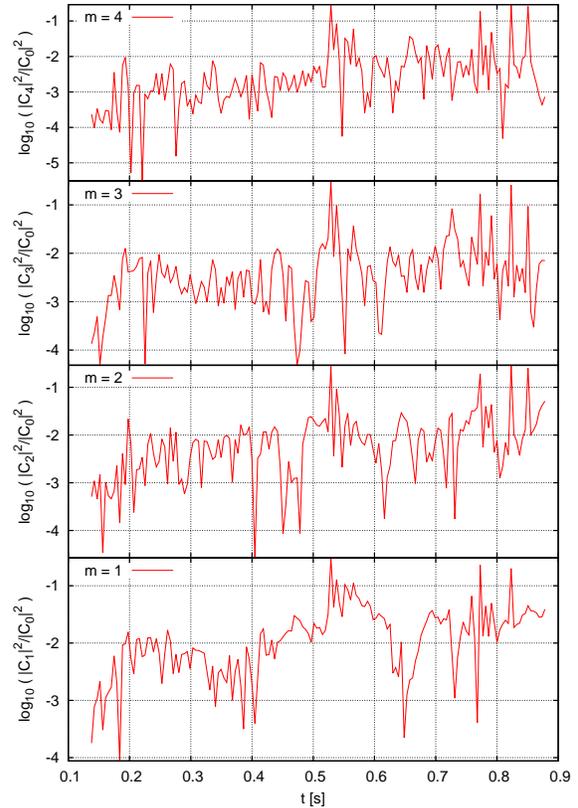}
  \caption{Relative power $|c_{m}|^{2}=|C_{m}|^{2}/|C_{0}|^{2}$ ($m=1,2,3,4$) for the azimuthal mass distribution $\Phi_{M}$ Fourier transform of model $\beta0.134$. Mode $m=1$  is the strongest at times $0.2\lesssim t\lesssim0.3$~s, where an strong variation is shown in Figure 28. A  structure formation event occurs at $t\simeq0.52$~s, when all modes reach $|c_{m}|^{2}\gtrsim0.1$.}
  \label{fig:Fourierb0134}
\end{center}
\end{figure}

\begin{figure}
\begin{center}
 \includegraphics[width=3.48in]{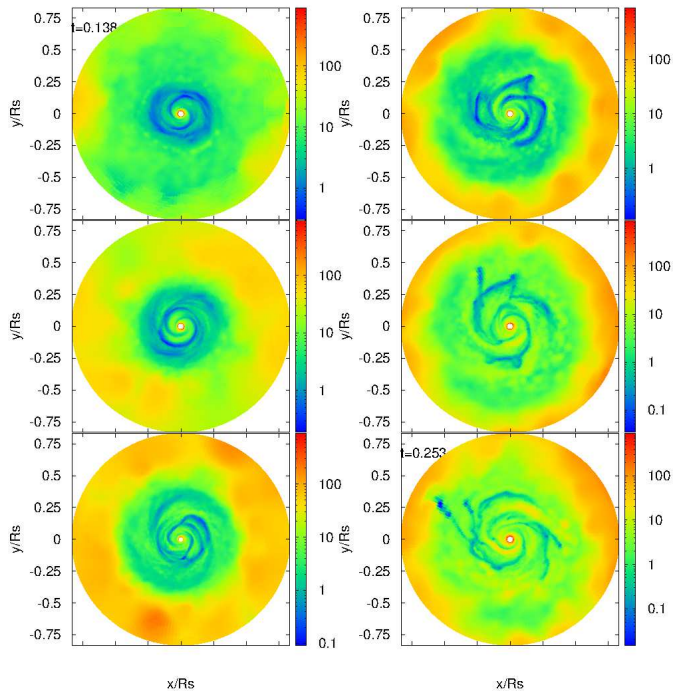}
  \caption{Evolution of the Toomre parameter $Q$ for model $\beta0.134$ at times $0.13\mbox{ s}\lesssim t\lesssim0.25\mbox{ s}$ . The spiral structures formad at very early times are quickly disrupted into clumps.}
  \label{fig:Toomre1b0134}
\end{center}
\end{figure}

The cooling time used in this model was so short that the envelope was significantly cooled
before $t\simeq t_{dyn}$. By looking at the Fourier modes amplitude $|c_{m}|^{2}$ in Figure~\ref{fig:Fourierb0134}, we can see that modes $m=1, 2$  are quite intense ($|c_{m}|^{2}>10^{-2}$) before $t\simeq0.2$ s when the accretion disk has just been formed. This suggests that we should see spiral structure at times as early as $0.15$ s. This can be seen in Figure~\ref{fig:Toomre1b0134} where  the Toomre parameter $Q$ evolution is plotted for times $0.13 \mbox{ s} \lesssim t\lesssim 0.25\mbox{ s}$, where we can observe the presence of spiral arms that form and rapidly dissipate. Before $t\simeq0.2$ s there is an increase in all Fourier modes which seems to be related to the intense variation seen in Figure~\ref{fig:MLb0134} in both $\dot{M}$ and $L_{c}$ at those times.

It is difficult to associate the intense and rapid increase shown in $\dot{M}$ and $L_{c}$ at $t\simeq0.28$~s in Figure~\ref{fig:MLb0134} with the spiral structure formation event seen in Figure~\ref{fig:Fourierb0134} given that all modes oscillate between $10^{-3}\lesssim|c_{m}|^{2}\lesssim10^{-2}$ right before $t\simeq0.3$ s. The same happens with the increase shown in Figure~\ref{fig:MLb0134} at $t\simeq0.35$ s, and which lasts for $\sim0.1$ s. Nevertheless, the increase shown on Figure~\ref{fig:MLb0134} at $t\gtrsim0.5$ s can be associated with an intense increase on all Fourier modes at the same time from Figure~\ref{fig:Fourierb0134}. At $t\simeq0.525$ s all Fourier modes reach a maximum value $\gtrsim10^{-1}$, which is not attained by less efficiently cooled models. This means that the mass ratio between the spiral structures and the disk's mass is greater than in previous models, which can be due to the fact that the BH has accreted a significantly greater amount of mass from the disk (more than twice the mass accreted with respect to less efficiently cooled models). 

\begin{figure}
\begin{center}
 \includegraphics[width=3.48in]{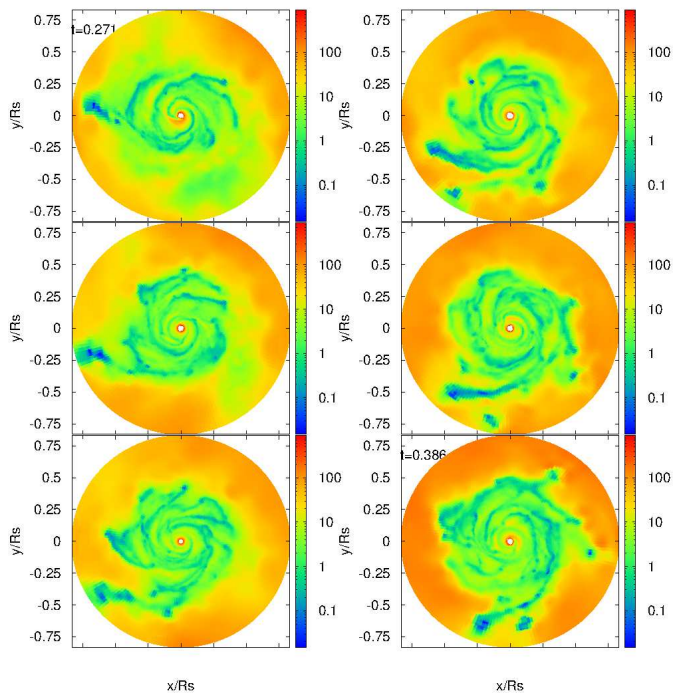}
  \caption{Evolution of the Toomre parameter $Q$ for model $\beta0.134$ at times $0.27\mbox{ s}\lesssim t\lesssim0.38\mbox{ s}$. A large number of clumps with low  $Q$ values are seen orbiting at variosu distances from the BH. All gas with $Q \leq 1.5\times 10^{-2}$ is plotted as the darkest blue color, even thought it can reach values as low as $Q\sim10^{-3}$.}
  \label{fig:Toomre2b0134}
\end{center}
\end{figure}

\begin{figure}
\begin{center}
 \includegraphics[width=3.48in]{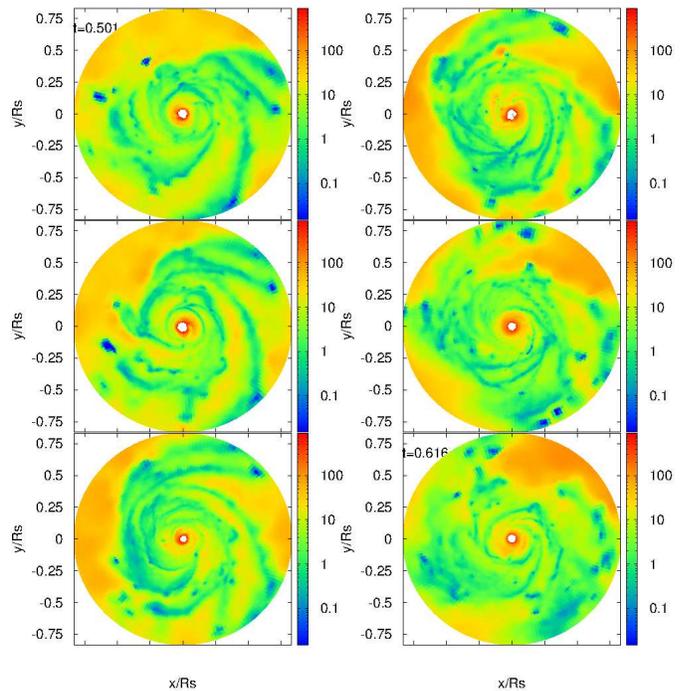}
  \caption{Evolution of the Toomre parameter $Q$ for the $\beta0.134$ model at times  $0.50\mbox{ s}\lesssim t\lesssim0.61\mbox{ s}$. All gas with $Q \leq 1.5\times 10^{-2}$ is plotted as the darkest blue color, even thought it can reach values as low as $Q\sim10^{-3}$.}
  \label{fig:Toomre3b0134}
\end{center}
\end{figure}

Figures~\ref{fig:Toomre2b0134} and \ref{fig:Toomre3b0134} show the evolution of the Toomre parameter $Q$ at times $0.27\mbox{ s}\lesssim t\lesssim0.38\mbox{ s}$, and $0.50\mbox{ s}\lesssim t\lesssim0.61\mbox{ s}$ respectively. As can be seen from both figures, $Q$ reaches higher values within the disk, probably because the surface density is considerably lower after the rapid BH mass accretion. Nevertheless, all regions ranging from green to blue have $Q\lesssim 1$ and it is clear that there are intense spiral structures and clumps forming in the disk. These clumps can be observed as early as $t\simeq0.2$ s. In order to better appreciate the presence of instabilities in the disk (regions with $Q\lesssim1$), we fixed a lower limit on the Toomre parameter at 0.02 so that all regions with $Q\lesssim1$ will have a color between blue and green ($Q$ reaches values as low as $\sim0.002$).

This is the model with one of the highest cooling efficiencies that we have applied to our simplified system, and its results should be taken with care when relating them to an actual collapsar model due to the fact that the envelope was significantly cooled down before reaching the BH. Nevertheless, the intense structure formation observed is something we would expect to see in the innermost part of the disk of a collapsar, given that the neutrino cooling efficiency is a very steep function of temperature.

\section{Summary and Conclusions}\label{conclusions}

We have presented a thorough 3D numerical study of the accretion of infalling envelopes onto black holes, using an simplified equation of state for the gas, considering that the ideal gas contribution from free nuclei dominates over radiation and a hot e$^{\pm}$ gas. We have further included a simplified prescription for cooling  based on realistic emission rates expected at the given densities and temperatures within a collapsar, with which we can study scenarios ranging from adiabatic to isothermal, with the aim of identifying and characterizing the morphology and the variations in structure induced within the disk as a result of the energy losses to neutrinos. With these caveats, which we discuss further below, we find global features which are likely to be present in real collapsing stellar cores and are relevant for the dynamics and energy release leading to the production of cosmological GRBs from massive stars. 

Here we summarize our main results:

$\bullet$
The most important parameter governing the energy release in the collapse of a stellar core in the context of a collapsar is the rotation rate. If it is too low, the gas will flow essentially in radial fashion, and accrete onto the central black hole releasing very little of its energy (akin to Bondi flow). If it is too high, the disk will form at a radius that may be too large for efficient cooling to kick in (recall the sensitive dependence on neutrino emissivity on temperature), and the accretion efficiency may be too low to provide sufficient energy. It is the combination of placing shocked matter in centrifugal support as close to the BH as possible (and hence deep in the potential well), but not allowing it to fall in, that is critical for a successful event. In 2D, this has been characterized before (Lee \& Ramirez-Ruiz 2006, L\'{o}pez-C\'{a}mara et al. 2009, 2010), and it is plainly important in 3D as well. In this sense, modeling the black hole through the pseudo potential of Paczynski \& Wiita (
1980) is the most important ingredient as far as gravitational effects are concerned, which is why we have adopted it in the present work. Of course, considering a rotating BH may change the quantitative results somewhat, but not the qualitative nature of this conclusion.

$\bullet$ Increasing the cooling efficiency induces more profuse and intense structure formation which in
turn produces strong variations both in the mass accretion rate and the energy losses. The duration and intensity of these variations is related with the intensity and lifetime of the collapsed structures, whether they are spiral arms or clumps. Structures with a Toomre parameter $Q\ll1$ and high relative power $|c_{m}|^{2}$ contain considerable amounts of gas within compact regions, and can therefore be accreted quickly, leading to large variations in $\dot{M}$ and $L_{c}$. The frequency of these variations (and structure formation) increases
with rising cooling efficiency, as observed in the Fourier power spectra of  $\dot{M}$ and $L_{c}$'s.

$\bullet$ The spiral structures that form are clearly transient in nature, forming and disappearing within a few orbital periods. Their relative power is particularly strong in the lowest azimuthal modes, $m=1,2$ at modest cooling, when $t_{cool}\simeq t_{disk}$ and rapidly spreads to higher modes, $m=3,4$ for increasing cooling efficiency, when $t_{cool}\ll t_{disk}$. 

$\bullet$ The spiral structure and gas clumps in the disk not only induce intense variations (with different durations) in both $\dot{M}$ and $L_{c}$, but are also significant enough to break the symmetry in gravitational interaction between the disk and the black hole. This can be seen in Figures~\ref{Fou_1} and \ref{Fou_2} where the Fourier transforms of the radial momentum component $P_{r}$ of the disk are shown. As cooling becomes more efficient, greater power is seen at higher frequencies (shorter time scales) in $P_{r}$. Such characteristic frequencies are easily noticeable just by looking the time series of the components of momentum $P_{x}$ and $P_{y}$ separately, where one can identify a ``periodic'' oscillation of the accretion disk. By forcing the BH to remain fixed at the origin during the simulations, we cannot here give a full account of this effect, but clearly these oscillations could have important consequences for the overall dynamics of the flow and are further discussed below.

$\bullet$ The integrated energy losses obtained range from 1 to 2000 foe ($10^{51}$ to $10^{54}$ erg) for adiabatic to isothermal models. Now, this is only what would count as "neutrino" energy release. The integrated accretion energy $E_{acc}=\int \dot{M} c^{2} dt$ ranges from $10^{54}$ to $3\times10^{54}$ erg. Both of these indicate that as expected, the available power to drive a GRB is present in the system. How it eventually manages to do so is a different matter, but, more to the point here, it is clear that the time variations imprinted on the outflow are dependent on the initial mechanism driving it. This is in way akin to the argument presented in Carballido \& Lee (2011), where temporal variability in local shearing box simulations for different neutrino cooling prescriptions, when integrated over the entire disk on a large scale, produces a different and characteristic power spectrum which is the underlying shape upon which further processes are superimposed, 
each adding its typical signature layer of behavior. 

$\bullet$ 
The formation and destruction of accretion structures, in particular arms and clumps, is not only transient, but recurrent in an orderly fashion. Inspection of Figures~\ref{Mdot} and \ref{Lum}, where the mass accretion rates and energy release are plotted, shows that for greater efficiency, the interval between enhancements in $\dot{M}$ and $L_{c}$ is smaller. The disk is draining of matter in the process as well, of course, and so each subsequent episode is of lesser intensity. But the trend appears to be clear: as structure forms, the disruption of azimuthal symmetry allows for a degree of angular momentum transport through the dense spiral arms (or even clumps), leading to an accretion episode. Having suppressed this structure, a second cooling interval must elapse before new condensations form and allow the process to repeat, indicating a  correlation between the time elapsed between accretion (or luminosity) spikes and the cooling time itself. For decreasing cooling times the disk is being depleted ever 
faster and so 
the trend becomes increasingly difficult to see (note the accretion rate and luminosity are plotted in a logarithmic scale).


$\bullet$
Given our cooling implementation, Eq.~(\ref{cooling}), the  internal energy per unit mass of the accretion disk $u_{M}= U / M_{disk}$ is directly related to the energy loss rate $L$ by:
\begin{equation}
 u_{M} = \frac{U}{M_{disk}} = \frac{1}{M_{disk}}\left(\sum_{j} u_{i,j}\right) = \frac{t_{cool} L}{M_{disk}},
\end{equation}
where $u_{i,j}$ is the internal energy of the particle $j$ and $M_{disk}$ is the mass of the accretion disk at time $t$. Thus, $u_{mass}$ contains information on both the energy loss rate and the accretion rate (determining the disk's remaining mass). The evolution of $u_{mass}$ as a function of the normalized time $t/t_{cool}$, shows some intrinsic properties of this particular cooling scenario. 

\begin{figure}
\begin{center}
 \centering
 \includegraphics[width=2.40in,angle=-90]{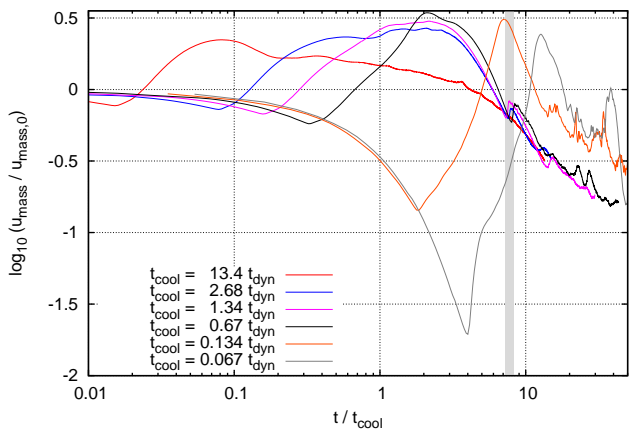}
  \caption{Internal energy per unit mass $u_{M}$ as a function of $t/t_{cool}$. The grey shaded region from $7.2 \leq t/t_{cool}\leq 8.2$ shows the times at which models with $t_{cool} \gtrsim t_{dyn}$ show the first structure formation event represented as an increase in $u_{M}$ (except for model $\beta13.4$).}
   \label{Umass}
\end{center}
\end{figure}

Figure \ref{Umass} shows the normalized internal energy per unit mass $u_{M}/u_{M,0}$ ($u_{M,0}= u_{M}(t=0)$) as a function of $t / t_{cool}$ for all simulations with our cooling prescription, Eq.~\ref{cooling}. All models show the initial decrease in $u_{M}$ due to the cooling and collapse of the envelope, followed by an intense increase from  the outward shock produced by the heated material reaching its centrifugal barrier near the black hole. Both of these events take place at times $t \lesssim t_{dyn}$ and therefore, models with $t_{cool} \gtrsim t_{dyn}$ will start accreting and will have formed an accretion disk at times $t \lesssim t_{cool}$. Meanwhile models with $t_{cool} << t_{dyn}$ will form an accretion disk only after several cooling times have elapsed. Models displaying such scenarios are intrinsically different, and hence we do not expect to observe similar behavior when comparing them. Models with $t_{cool} \gtrsim t_{dyn}$ display an increase in $u_{M}$ at a time $7.2 > t / t_{cool} > 8.2$, 
which corresponds to the structure formation event previously noted in models $\beta 2.68$, $\beta 1.34$ and $\beta 0.67$. The exception to this trend is model $\beta 13.4$, which shows no significant increase in $u_{M}$ and no structure formation event at such times. 

\begin{figure}
\begin{center}
 \centering
 \includegraphics[width=2.40in,angle=-90]{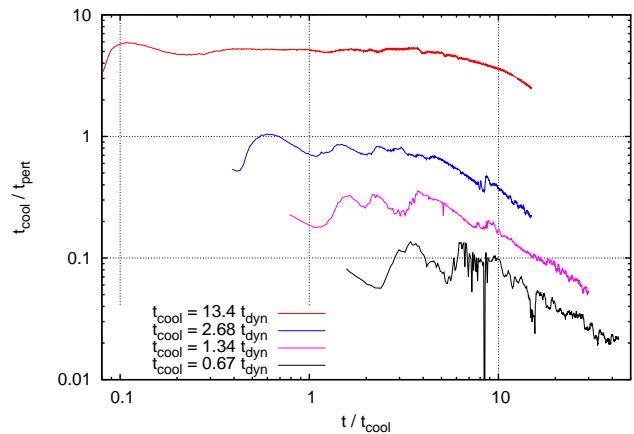}
  \caption{Time evolution of the ratio between $t_{cool}$ and $t_{pert}$ for models $\beta13.4$, $\beta2.68$, $\beta1.34$ and $\beta0.67$. Models with efficient cooling have ratios $\lesssim 1$ and are therefore, able to have structure formation events.}
   \label{Cs_t}
\end{center}
\end{figure}

The exception can be explained by considering the minimum azimuthally averaged sound speed in the disk $c_{s,min} = \min (c_{s}(R),(0.05 < R/R_{s} < 1 )$, from which we can estimate an upper limit for the time, $t_{pert} = R_{s} / c_{s,min}$,  it would take a pressure perturbation to transfer its information throughout a disk of size $R_{s}$. In particular, this perturbation could be induced  precisely by cooling. Thus, if the cooling time scale is smaller than this perturbation time scale $t_{pert}$, there is a region within the disk where any significant drop in pressure (brought about by a reduction in the internal energy), could not be immediately compensated by hydrodynamical processes and it could experience a collapse if the pressure drop is strong enough. On the other hand, if the cooling time scale is significantly higher than $t_{pert}$, any drop in pressure will be quickly accounted for, and softened, by hydrodynamical processes before the gas gets cool enough to undergo collapse. 
This is the case in model $\beta 13.4$ which, at all times, satisfies $t_{cool} > 2 t_{pert}$,  as seen on Figure \ref{Cs_t} where the evolution of $t_{cool} / t_{pert}$ vs $t / t_{cool}$ is plotted. Hence, we do not expect to see any structure formation event from this model as long as the the condition $t_{cool} / t_{pert} \lesssim 1$ is not satisfied.

Clearly there is also room for improvement in what we have presented here, and we detail some of these issues here:

$\bullet$ First, a more detailed implementation of neutrino cooling is likely to lead to more realistic results. Particularly, this refers to the fact that we have assumed a uniform cooling efficiency throughout the flow, $\beta$. As the temperature dependence of cooling is significant, the outer regions will not emit as copiously as the inner disk, leading to different behavior. We believe this is one of the two most important points which would require addressing. 

$\bullet$ We have alluded to the second above already, namely, that the gravitational interaction between the disk and the black hole does not always allow one to assume that the latter is always lying at the origin. Rather, it will oscillate with the disk, as in a binary. This has two important consequences: first, the accretion rate will be modified somewhat, as clumps and spiral arms will not be disrupted/accreted in the same way, and second, the site of accretion, and potential jet driving through the envelope, will be in continuous motion within the disk and stellar envelope. A quantitative analysis of this is effect is clearly necessary, and it may have far-reaching consequences, as the common assumption has been that the energy deposition driving a relativistic outflow leading to breakout sits motionless at the center of the star. 

$\bullet$
Although the PW pseudo potential, as mentioned above, captures the most essentially feature of General Relativity for accretion purposes, a full relativistic treatment of the problem is desirable. In particular, not only the effects of a spinning (Kerr) black hole, but of the field produced by the flow itself, which, as we have seen, can in some cases produce a significant pull on the central object. 

$\bullet$
Finally, a proper treatment of neutrinos should consider the effects of neutrino transport and energy deposition. Since the emitted neutrinos could be able to deposit a significant amount of energy in the gas at high density regions ($\rho\gtrsim10^{11}\mbox{ g cm}^{-3}$) due to inelastic scattering off free nucleons and $\alpha$ particles, and because of neutrino opacity effects even in the elastic scattering regime, some of the structure formation we see may be inhibited, or limited, preventing the formation of the densest features. 

The rich behavior seen here, and characterized for the first time in 3D, clearly shows that using only 2D studies of the collapse of stellar cores is insufficient to explain all of the behavior and variability, even qualitatively, that is likely to occur in such systems. As we already know that only a small fraction of collapsing massive stars will produce a GRB, it is relevant in the sense of identifying precisely which conditions will lead to one in the evolution of the star. Preliminary work on the first two topics, in particular the disk-BH interaction, leads us to believe that these may be significant for the behavior and evolution of collapsar disks, and they will be the subject of future work. 

\textbf{Acknowledgements}

We gratefully acknowledge helpful conversations with Dany Page, Alejandro Raga and Enrico Ramirez-Ruiz. This work was supported in part by CONACyT (101958) and Direcci\'{o}n General de Estudios de Posgrado-UNAM. A Batta acknoledges support from a CONACyT graduate fellowship.

\label{lastpage}

\end{document}